\pdfoutput=1
\documentclass[11pt]{article}

\usepackage[final]{acl}

\usepackage{times}
\usepackage{latexsym}

\usepackage[T1]{fontenc}
\usepackage[utf8]{inputenc}

\usepackage{microtype}

\usepackage{inconsolata}

\usepackage{graphicx}
\usepackage{cleveref}

\usepackage{longtable}

\title{Findings of the BEA 2025 Shared Task on \\Pedagogical Ability Assessment of AI-powered Tutors}

\author{Ekaterina Kochmar$^1$, Kaushal Kumar Maurya$^1$, Kseniia Petukhova$^1$, \\ {\bf KV Aditya Srivatsa$^1$, Anaïs Tack$^2$, Justin Vasselli$^3$} \\
  $^1$Mohamed bin Zayed University of Artificial Intelligence (MBZUAI), \\ $^2$KU Leuven, $^3$Nara Institute of Science and Technology \\
}

\begin{document}
\maketitle
\begin{abstract}
This shared task has aimed to assess pedagogical abilities of AI tutors powered by large language models (LLMs), focusing on evaluating the quality of tutor responses aimed at student's mistake remediation within educational dialogues. The task consisted of five tracks designed to automatically evaluate the AI tutor's performance across key dimensions of \textit{mistake identification}, \textit{precise location of the mistake}, \textit{providing guidance}, and \textit{feedback actionability}, grounded in learning science principles that define good and effective tutor responses, as well as the track focusing on {\em detection of the tutor identity}. 
%Given real tutor and student conversation history and current tutor response from state-of-the-art LLMs and humans, participants were tasked with developing automated evaluation models capable of predicting the response quality according to the pedagogical aspects. 
The task attracted over 50 international teams across all tracks. The submitted models were evaluated against gold-standard human annotations, and the results, while promising, show that there is still significant room for improvement in this domain: the best results for the four pedagogical ability assessment tracks range between macro F1 scores of 58.34 (for {\em providing guidance}) and 71.81 (for {\em mistake identification}) on three-class problems, with the best F1 score in the {\em tutor identification} track reaching 96.98 on a 9-class task. In this paper, we overview the main findings of the shared task, discuss the approaches taken by the teams, and analyze their performance. All resources associated with this task are made publicly available to support future research in this critical domain.\footnote{\url{https://github.com/kaushal0494/UnifyingAITutorEvaluation/tree/main/BEA_Shared_Task_2025_Datasets}}
\end{abstract}

\section{Introduction and Motivation}

Conversational agents offer promising opportunities for education as they can fulfill various roles (e.g., intelligent tutors and service-oriented assistants) and pursue different objectives (e.g., improving student skills and increasing instructional efficiency)~\cite{wollny2021we}, among which serving as an AI tutor is one of the most prevalent tasks~\cite{tack2023bea}. Recent advances in the development of large language models (LLMs) provide our field with promising ways of building AI-based conversational tutors, which can generate human-sounding dialogues on the fly. The key question posed in previous research~\cite{tack2022ai,tack2023bea}, however, still holds: {\em How can we test whether state-of-the-art generative models are good AI teachers, capable of replying to a student in an educational dialogue?}

% Paragraph 01: A general introduction to the research problem within the domain, outlining the scope and significance of the research space.

Evaluating dialogue systems in general presents a significant challenge. While human evaluation is still considered the most reliable method for assessing dialogue quality, its high cost and lack of reproducibility have led to the adaptation of both reference-based and reference-free automatic metrics, originally used in machine translation and summary evaluation, for dialogue evaluation~\citep{lin2004rouge,popovic2017chrf++,post2018call,gao2020dialogue,liu2023g}. When it comes to Intelligent Tutoring Systems (ITSs), which also function as dialogue systems with the specific role of acting as tutors, these general metrics are insufficient. In the educational context, we need to assess complex pedagogical aspects and abilities of such systems, ensuring that they provide students with sufficient, helpful, and factually correct guidance and do not simply reveal answers when the student makes a mistake, among other aspects. Therefore, developing automatic metrics to evaluate these nuanced aspects is essential for creating effective and helpful tutoring systems.

% Paragraph 02: Contextualization of the problem in relation to previous work, identifying key limitations, such as the absence of a standardized evaluation taxonomy.
Due to the lack of a standardized evaluation taxonomy, previous work has used different criteria for evaluation. For example, ~\citet{tack2022ai} and \citet{tack2023bea} evaluated models in terms of whether they {\em speak like a teacher}, {\em understand a student}, and {\em help a student}, while in~\citet{macina2023mathdial}, responses of models playing roles of tutors were evaluated by human annotators using {\em coherence}, {\em correctness}, and {\em equitable tutoring}. At the same time, ~\citet{wang2024bridging} assessed {\em usefulness}, {\em care}, and {\em human-likeness}, and \citet{daheim2024stepwise} used {\em targetedness}, {\em correctness}, and {\em actionability} of a tutor response as quality evaluation criteria. Such lack of standardization makes it difficult to compare different systems, and, therefore, defining evaluation criteria and developing automatic metrics for them is a crucial task for advancing the field, which we have aimed to address in this task.

% Paragraph 03: A brief introduction to evaluation rubrics, with an emphasis on tantativly three critical dimensions: (a) Has spotted the mistake?, (b) Has guidance?, and (c) Has actionability?, aligning these aspects with established principles from learning sciences.

% Paragraph 04: brief about the relevance to the BEA community

\section{Task Description and Goals}
\label{sec:descript}
% Clear goals for the tasks: (1) Subtask-1, (2) Subtask-2 and (3) Subtask-3  

Following the successful BEA 2023 Shared Task on {\em Generating AI Teacher Responses in Educational Dialogues}~\cite{tack2023bea}, we revisit the question of quality assessment of the tutor responses generated with the AI models (specifically, LLMs) in the context of educational dialogues. We believe that (1) the topic is timely and important; (2) LLMs have significantly advanced in the past couple of years, making it important to revisit this topic after the competition run in 2023; and (3) there is a need to establish a pedagogically motivated benchmark for this task. In contrast to the BEA 2023 Shared Task, our focus is not on the {\em generation of educational dialogues} using state-of-the-art LLMs, but rather on {\bf comprehensive evaluation of AI-tutor responses using a set of pedagogically motivated metrics}.

In this shared task, we have focused on educational dialogues between a student and a tutor in the mathematical domain. Specifically, the conversations are grounded in student mistakes or confusion, where the AI tutor aims to remediate such mistakes or confusion. Each dialogue in the datasets provided in this shared task includes: (i) the context consisting of several prior turns from both the tutor and the student; (ii) the last utterance(s) from the student containing a mistake; and (iii) a set of possible responses to the last student's utterance(s) from a range of LLM-based tutors and, where available, human tutors, aimed at mistake remediation. The dialogues (parts i-ii) are extracted from two popular datasets of educational dialogues in the mathematical domain – {\tt MathDial}~\cite{macina2023mathdial} and {\tt Bridge}~\cite{wang2024bridging}, while the LLM-based tutor responses are generated by the shared task organizers using a set of state-of-the-art LLMs of various sizes and capabilities, including: {\tt GPT-4} \cite{achiam2023gpt}, {\tt Gemini} \cite{reid2024gemini}, {\tt Sonnet} \cite{TheC3}, {\tt Mistral} \cite{jiang2023mistral}, {\tt Llama-3.1-8B} and {\tt Llama-3.1-405B} \cite{dubey2024llama}, and {\tt Phi-3}~\cite{abdin2024phi}. To avoid any biases, the tutor responses in the data have been shuffled. In addition to the responses themselves, the dataset contains annotation of their quality along several pedagogically motivated dimensions defined in~\citet{maurya-etal-2025-unifying}. Below, we are reiterating the definitions for these dimensions from~\citet{maurya-etal-2025-unifying} for completeness:%\vspace{-0.5em}

\begin{itemize}
    \item {\em Mistake identification}: Since all dialogues in the dataset contain a mistake made by the student, we expect a good quality response from the tutor to include relevant and clear mistake identification. This aligns with {\em student understanding} defined in \citet{tack2022ai} and {\em correctness} in the schemata of \citet{macina2023mathdial} and \citet{daheim2024stepwise}.%\vspace{-0.5em}
    \item {\em Mistake location}: In addition to notifying the student about the committed error, a good tutor response should also point to its location in the answer and explain what the error is to help the student remediate it. This corresponds to {\em targetedness} in \citet{daheim2024stepwise}.%\vspace{-0.5em}
    \item {\em Providing guidance}: Ideally, a good tutor should not reveal the answer immediately and instead should provide the student with relevant and helpful guidance, consisting, for example, of a hint, an explanation, or a supporting question. This aspect is related to {\em helping a student} in \citet{tack2022ai} and {\em usefulness} in \citet{wang2024bridging}.%\vspace{-0.5em}
    \item {\em Actionability}: Finally, once the guidance is provided to a student, a good tutor response should make it clear to the student what they are supposed to do next. I.e., the tutor's response should not be vague, unclear, or a conversation stopper. This aspect directly corresponds to {\em actionability} in \citet{daheim2024stepwise}.%\vspace{-0.5em}
\end{itemize}

%learning science principles mapping:
% encourage active learning -> Revealing of the answer, Providing guidance
% manage cognitive load ->  Providing guidance
% deepen metacognition ->  Providing guidance, 
% motivate and stimulate curiosity ->  Actionability
% adapt to learners’ goals and needs -> Mistake identification, Mistake location

Moreover, the proposed evaluation schema aligns with the core pedagogical principles derived from learning sciences. Specifically, the tutor should: (1) \textit{encourage active learning} \cite{chi2014icap, oakley2021uncommon} by not directly revealing the correct answer, (2) \textit{adapt to learners' goals and needs} \cite{king2017reimagining} through accurate mistake identification and exact location pointing, (3) \textit{manage cognitive load}  \cite{mayer2002multimedia} and enhance \textit{metacognitive skills} \cite{dehaene2020we, cohen2021metacognitive} by providing appropriate guidance, and (4) foster \textit{motivation and stimulate curiosity} \cite{keller1987development, patall2008effects} by offering clear and actionable steps to the student. Thus, the schema adopted from~\citet{maurya-etal-2025-unifying} covers all the relevant aspects of a good tutor response proposed in previous work~\cite{tack2022ai,macina2023mathdial,wang2024bridging,daheim2024stepwise}, while also being supported by the learning science principles. We do not explicitly include such aspects as {\em speak like a teacher}~\cite{tack2022ai}, as we believe that a tutor that identifies student's mistakes, points to them accurately, and can explain them to a student in an actionable way {\em does} speak like a teacher. We also do not explicitly cover {\em human-likeness}~\cite{wang2024bridging} as, based on our preliminary analysis, state-of-the-art LLMs are capable of producing overwhelmingly human-like responses. 

\begin{figure}
    \centering
    \includegraphics[width=1\linewidth]{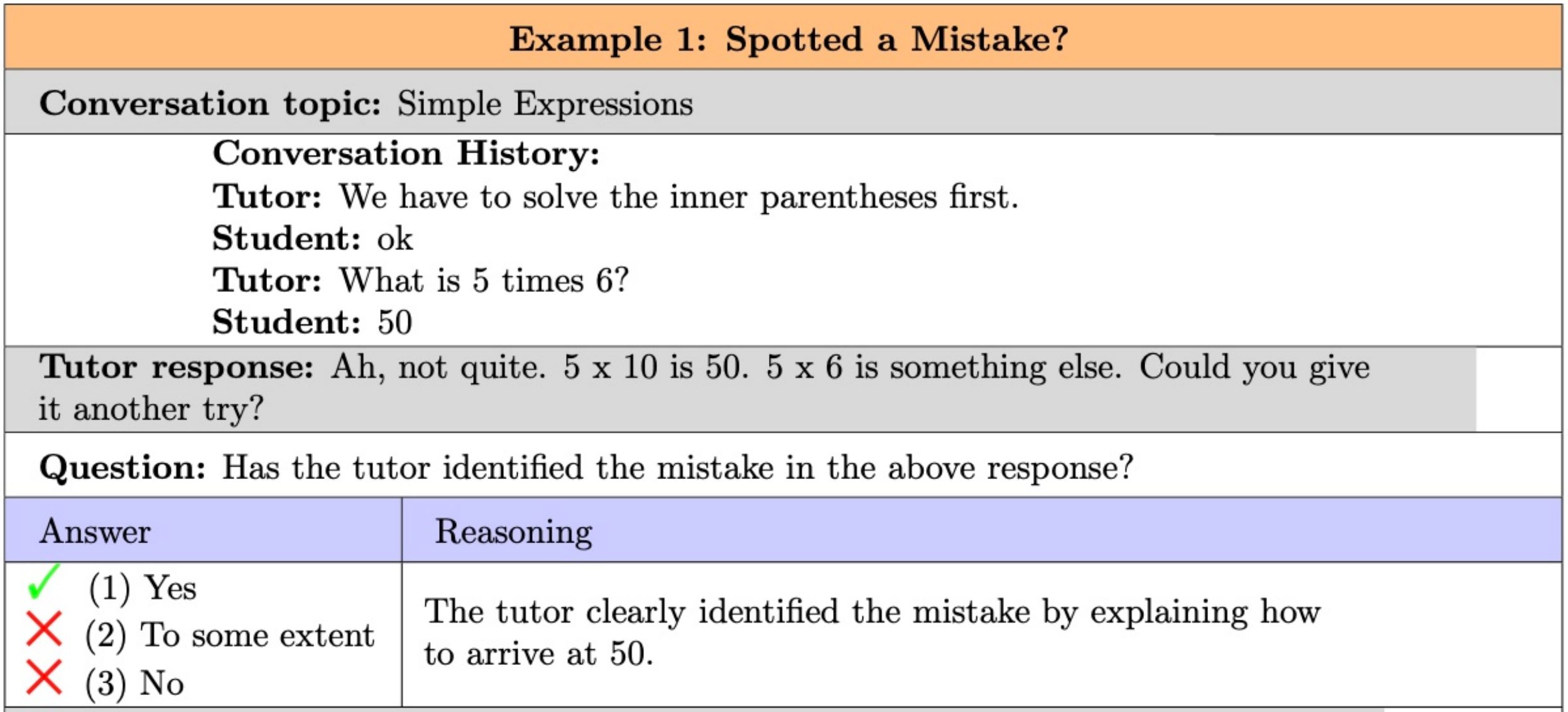}
    \vspace{-0.3cm}
    \caption{An example on mistake identification from~\citet{maurya-etal-2025-unifying}}
    \vspace{-0.3cm}
    \label{fig:example}
\end{figure}

All aspects are annotated on a 3-point scale, where "No" denotes that the particular aspect of the tutor response is {\em bad} (e.g., the mistake is not identified at all), "Yes" denotes that it is {\em good} (e.g., the mistake is identified clearly and correctly), and "To some extent" denotes that the quality of the response, according to the particular aspect, is medium (e.g., there are clarity issues with the mistake identification).
%TODO: Check correspondences
%~\citet{tack2022ai}: {\em speak like a teacher}, {\em understand a student}, and {\em help a student}
%while in~\citet{macina2023mathdial}: {\em coherence}, {\em correctness}, and {\em equitable tutoring}
%~\citet{wang2024bridging}: {\em usefulness}, {\em care}, and {\em human-likeness} 
%\citet{daheim2024stepwise}: {\em targetedness}, {\em correctness}, and {\em actionability} 
Figure \ref{fig:example} provides an example of the annotation for the {\em mistake identification} aspect.

\section{Shared Task Structure}
% A description of the shared task structure – how many tracks it will consist of, whether there will be closed and/or open data tracks, whether training data will be provided by the organizers or whether already existing datasets will be used, etc.

This shared task consisted of two major phases:\vspace{-0.5em}

\begin{itemize}
    \item {\bf Development phase}: In the development phase, we released annotated tutor responses for 300 dialogues extracted from the {\tt MathDial} and {\tt Bridge} datasets (approximately 75\% examples from {\tt MathDial} and 25\% examples from {\tt Bridge}). For each dialogue, responses from 7 LLM-based tutors (see Section \ref{sec:descript} for more details) as well as expert (for both datasets) and novice (provided in the {\tt Bridge} dataset only) tutor responses were released together with the annotations for 4 pedagogical aspects following the scheme and guidelines proposed in \citet{maurya-etal-2025-unifying}. This sums up to a total of 2,476 tutor responses. During the development phase, participating teams could build their systems aiming to predict the quality values for any or all of the pedagogical aspects.\vspace{-0.5em}
    \item {\bf Test phase}: In the test phase of the competition, we released an additional set of 191 dialogues extracted from the {\tt MathDial} and {\tt Bridge} datasets, following the distribution in the development set, together with the tutor responses (1,547 in total), but the annotations were not provided for this data. The participating teams were asked to run their systems and submit their predictions, which were then evaluated using the shared task official metrics (see Section \ref{sec:evaluation}).\footnote{Development and test sets are available at \url{https://github.com/kaushal0494/UnifyingAITutorEvaluation/tree/main/BEA_Shared_Task_2025_Datasets}}\vspace{-0.5em}
\end{itemize}

In addition, the task included the {\em 5th track} on the tutor identity identification, aimed at automated detection of which model or human tutor an anonymous response in the test data originated from. This sub-task was inspired by our observations that various AI tutors have very specific tutoring and linguistic styles~\cite{maurya-etal-2025-unifying}.

The task used {\em open-data and model strategy}: as there were no explicit training phase, the teams were allowed to use any external data in addition to the released annotated dialogues during the development phase, as well as build traditional machine learning as well as large language model-based solutions. 

The test phase of the task was hosted on the CodaBench platform, with a separate track for {\em (1) Mistake Identification},\footnote{\url{https://www.codabench.org/competitions/7195/}} {\em (2) Mistake Location},\footnote{\url{https://www.codabench.org/competitions/7200/}}{\em (3) Providing Guidance},\footnote{\url{https://www.codabench.org/competitions/7202/}} {\em (4) Actionability},\footnote{\url{https://www.codabench.org/competitions/7203/}} and {\em (5) Tutor Identification}.\footnote{\url{https://www.codabench.org/competitions/7206/}} Each team was allowed up to 5 submissions in each track.

\section{Data Description}
%A detailed description of the data, its novelty and its preparedness for the task – i.e., current availability of the dataset(s) for the task, availability of the annotations if relevant, amount of data, its format, etc.
%Specification of whether the data will be made publically available to the scientific community after the task and whether there are any restrictions on its use and dissemination.

As described in Section \ref{sec:descript}, we used the data from two publicly available datasets – {\tt MathDial}~\cite{macina2023mathdial} and {\tt Bridge}~\cite{wang2024bridging}. Both datasets allow for adaptation, modification and (re)sharing without any restrictions:
{\tt MathDial}\footnote{\url{https://github.com/eth-nlped/mathdial}} is distributed under the Creative Commons Attribution-ShareAlike 4.0 International License, while the {\tt Bridge}\footnote{\url{https://github.com/rosewang2008/bridge}} dataset is licensed under the MIT license.

Annotations of the pedagogical aspects were provided by the organizing team following the scheme and guidelines established by~\citet{maurya-etal-2025-unifying}. Of the 300 development set dialogues, responses in 200 were doubly-annotated by four annotators, reaching an average Fleiss’ kappa of 0.65, which indicates substantial agreement and shows reliability of this task (see the description of the annotation experiment in ~\citet{maurya-etal-2025-unifying}). An additional set of tutor responses for further development and test set dialogues were annotated by the six shared task organizers using the same scheme and approach. A subset of 83 tutor responses in 10 dialogues were annotated by all six co-organizers, showing substantial agreement on the use of the scheme with Fleiss' kappa of 0.64. After this initial annotation round, co-organizers discussed and resolved disagreements before proceeding to annotate the rest of the data.

\section {Teams}

Over 50 teams participated in the shared task, with 11 teams submitting to all five tracks. The task attracted participation from all over the world, with teams from Asia (e.g., Bangladesh, China, India, Indonesia, Philippines, and South Korea), Australia, Europe (e.g., France, Germany, and Romania), the MENA region (e.g., Egypt, Lebanon, and the UAE), the North (e.g., USA and Canada) as well as South America (e.g., Chile and Uruguay) taking part in it. The submissions were distributed as reported in Table \ref{tab:overall}.\footnote{The official leaderboards can be found in Appendix \ref{app:leaderboards}.} The next section briefly summarizes the main trends in the approaches adopted by the teams, while more details can be found in the individual system reports submitted by 26 teams as well as in Section \ref{sec:evaluation}. 

\begin{table}[!htb]
    \centering
    \resizebox{0.3\textwidth}{!}{
    \begin{tabular}{|l|r|r|}
    \hline
        \textbf{Track} & \textbf{\# Submissions} & \textbf{\# Teams} \\ \hline\hline 
       Track 1 & 153 & 44 \\ \hline
       Track 2 & 86 & 32 \\ \hline
       Track 3 & 105 & 36 \\ \hline
       Track 4 & 87 & 30 \\ \hline
       Track 5 & 54 & 20 \\ \hline
    \end{tabular}}
    \caption{Number of submissions and participating teams in each track}
    \label{tab:overall}
\end{table}

%\begin{itemize}
%\item {\bf Track 1} attracted 153 submissions from 44 teams\vspace{-0.5em}
%\item {\bf Track 2} received 86 submissions from 32 teams\vspace{-0.5em}
%\item {\bf Track 3} got 105 submissions from 36 teams\vspace{-0.5em}
%\item {\bf Track 4} received 87 submissions from 30 teams\vspace{-0.5em}
%\item {\bf Track 5} attracted 54 submission from 20 teams\vspace{-0.5em}
%\end{itemize}

\subsection {Main Trends}

Based on the overall analysis of the approaches taken by the participating teams, we have identified the following major trends:
\begin{itemize}
    \item A few teams used {\em LLMs}, both commercial (\texttt{GPT-4o}~\cite{hurst2024gpt}, \texttt{Gemini}~\cite{reid2024gemini}, \texttt{Claude}~\cite{TheC3}) and open-source (\texttt{Mistral}~\cite{jiang2023mistral}, \texttt{LLaMa}~\cite{dubey2024llama}, \texttt{Qwen}~\cite{bai2023qwen}) extensively. Examples include teams {\tt BJTU}~\cite{fan-2025-shared}, {\tt BLCU-ICALL}~\cite{an-2025-shared}, {\tt NeuralNexus}~\cite{naeem-2025-shared}, {\tt Henry}~\cite{pit-2025-shared}, and {\tt LexiLogic}~\cite{bhattacharyya-2025-shared}, among others.
    \item {\em LoRA-based fine-tuning}~\cite{hu2022lora} has also been popular among the participants, including teams {\tt TutorMind}~\cite{dekmak-2025-shared}, {\tt Archaeology}~\cite{roșu-2025-shared}, {\tt Wonderland\_EDU@HKU}~\cite{deliang-2025-shared}, {\tt Averroes}~\cite{yasser-2025-shared}, and {\tt MSA}~\cite{hikal-2025-shared}.
    
    \item {\em Data augmentation and imbalance handling} were used, including methods like synthetic data generation by {\tt TutorMind}~\cite{dekmak-2025-shared} and {\tt Henry}~\cite{pit-2025-shared}, random downsampling by {\tt BJTU}~\cite{fan-2025-shared}, oversampling by {\tt Thapar Titans}~\cite{dadwal-2025-shared} and {\tt NLIP}~\cite{saha-2025-shared}, and class-weighted loss by {\tt Jinan Smart Education}~\cite{chen-2025-shared} and {\tt SYSUpporter}~\cite{chen-huang-2025-shared}.
    
    \item {\em Ensemble methods} were also applied: this included majority voting by {\tt Jinan Smart Education}~\cite{chen-2025-shared}, stacking by {\tt NLIP}~\cite{saha-2025-shared}, and disagreement-aware inference by {\tt MSA}~\cite{hikal-2025-shared}.                
    \item Finally, {\em hybrid and multi-stage architectures} were used, including integration of simpler models for initial prediction followed by escalation to more powerful LLM judges as in the approach by {\tt Emergent Wisdom}~\cite{jain-2025-shared}, or use of architectures that combine embeddings and classification models as in the dual-encoder setup used by {\tt Jinan Smart Education}~\cite{chen-2025-shared}.
\end{itemize}

\section{Evaluation, Results, and Summary of Approaches}
\label{sec:evaluation}
% A description of how the submissions will be evaluated.

Tracks 1-4 used {\bf macro F1 as the main metric}, with accuracy being the secondary metric. These were used in two settings:

\begin{itemize}
\item {\em Exact evaluation}: predictions submitted by the teams were evaluated for the exact prediction of the three classes (``Yes", ``To some extent", and ``No")\vspace{-0.5em}
\item {\em Lenient evaluation}: since for these dimensions tutor responses annotated as ``Yes" and ``To some extent" share a certain amount of qualitative value, we considered ``Yes" and ``To some extent" as a single class, and evaluated predictions under the 2-class setting (``Yes + To some extent" vs. ``No")\vspace{-0.5em}
\end{itemize}

Track 5 on {\em Tutor Identification} used {\bf macro F1 as its main metric}, and accuracy of the tutor identity prediction as its secondary metric, in an exact multi-class scenario without the {\em lenient} setting.

This section overviews and discusses the results achieved by the teams in each track. For the full leaderboards, see Appendix \ref{app:leaderboards}.

\subsection{Track 1: Mistake Identification}

Table \ref{tab:MI} presents the results of a majority-class baseline prediction model for the development ({\tt Dev maj.}) and test ({\tt Test maj.}) sets. Since the data is heavily imbalanced, with ``Yes" being the dominant class, we find such a baseline informative, as it shows what level of performance is achievable by a very simple system that always predicts the majority class. We report exact (strict) macro F1 ({\tt Ex. F1}) and accuracy ({\tt Ex. Acc}), as well as lenient F1 ({\tt Len. F1}) and accuracy ({\tt Len. Acc}).

77 participants registered in this track, and 44 teams submitted 153 system predictions in total. Table \ref{tab:MI} reports the best results achieved by the teams ({\tt Best test}) on all four metrics: exact F1 of $0.7181$, exact accuracy of $0.8798$, lenient F1 of $0.9185$, and lenient accuracy of $0.9541$. The winning team in this track, according to the main shared task metric (exact F1), as well as according to the secondary metric of lenient F1, is {\tt BJTU}~\cite{fan-2025-shared}.\footnote{The notation in brackets indicates the place according to the main (exact F1-based) ranking of the submission showing the best result for each individual metric.} The winners according to exact accuracy are {\tt TutorMind}~\cite{dekmak-2025-shared} and {\tt MSA}~\cite{hikal-2025-shared}, with {\tt TutorMind} scoring first in terms of lenient accuracy as well.

\begin{table}[!htb]
    \centering
    \resizebox{0.48\textwidth}{!}{
    \begin{tabular}{|l|l|l|l|l|}
    \hline
        \textbf{Category} & \textbf{Ex. F1} & \textbf{Ex. Acc} & \textbf{Len. F1} & \textbf{Len. Acc} \\ \hline\hline 
       Dev maj. & 0.2922 & 0.7803 & 0.4596 & 0.8506 \\ \hline \hline
       Test maj. & 0.2827 & 0.7363 & 0.4522 & 0.8255 \\ \hline
       Best test & 0.7181$^{(1)}$ & 0.8798$^{(12,13)}$ & 0.9185$^{(19)}$ & 0.9541$^{(35)}$ \\ \hline
    \end{tabular}}
    \caption{Results for {\em Track 1: Mistake Identification}}
    \label{tab:MI}
\end{table}

In this track, the 1st-place \texttt{BJTU} team used zero-shot prompting combined with dialogue-shuffling, random downsampling, and task-oriented prompt refinement~\cite{fan-2025-shared}. The 2nd-place \texttt{TutorMind} team fine-tuned \texttt{GPT-4o-mini} and \texttt{Mistral-7B} with LoRA and augmented their traing data synthetically, significantly improving model performance~\cite{dekmak-2025-shared}. \texttt{Averroes}, ranked 3rd, benchmarked multiple instruction-tuned models, demonstrating that compact, carefully tuned models could outperform larger ones~\cite{yasser-2025-shared}. The 4th-place \texttt{MSA} team used \texttt{Mathstral-7B} with LoRA and introduced disagreement-aware ensemble strategy~\cite{hikal-2025-shared}. Finally, the 5th-place \texttt{BD} team combined MPNet fine-tuning~\cite{song2020mpnet} with cross-validation and ensemble voting~\cite{rohan-2025-shared}.

\subsection{Track 2: Mistake Location}

Table \ref{tab:ML} presents the results of a majority-class baseline prediction model for the development and test sets, as well as the best results achieved by the participating teams on the test set. 

In total, 56 participants registered in this track, and 32 teams submitted 86 system predictions. Table \ref{tab:ML} reports the best results achieved by the teams ({\tt Best test}) on all four metrics: exact F1 of $0.5983$, exact accuracy of $0.7679$, lenient F1 of $0.8404$, and lenient accuracy of $0.8630$. The winning team in this track according to the main shared task metric (as well as exact and lenient accuracy) is {\tt BLCU-ICALL}~\cite{an-2025-shared}. The winner according to lenient F1 is {\tt K-NLPers}~\cite{park-2025-shared}.

\begin{table}[!htb]
    \centering
    \resizebox{0.48\textwidth}{!}{
    \begin{tabular}{|l|l|l|l|l|}
    \hline
        \textbf{Category} & \textbf{Ex. F1} & \textbf{Ex. Acc} & \textbf{Len. F1} & \textbf{Len. Acc} \\ \hline\hline 
       Dev maj. & 0.2560 & 0.6232 & 0.4159 & 0.7120 \\ \hline \hline
       Test maj. & 0.2450 & 0.5811 & 0.3974 & 0.6593 \\ \hline
       Best test & 0.5983$^{(1)}$ & 0.7679$^{(1)}$ & 0.8404$^{(5)}$ & 0.8630$^{(1)}$ \\ \hline
    \end{tabular}}
    \caption{Results for {\em Track 2: Mistake Location}}
    \label{tab:ML}
\end{table}

In this track, the 1st-place \texttt{BLCU-ICALL} team used a combination of in-context learning (ICL) with advanced prompting using the \texttt{Gemini-2.5-pro} model, supervised fine-tuning on large models like \texttt{Qwen2.5-32B}, and reinforcement learning from human feedback (RLHF)~\cite{an-2025-shared}. The 3rd-place \texttt{K-NLPers} used \texttt{GPT-4.1} combined with a specialized Multi-Perspective Reflective Evaluation approach, modeling internal deliberation among distinct reasoning perspectives~\cite{park-2025-shared}. The 5th-place team {\tt SG} used {\tt Gemma-3-27B-IT} in a two-step approach, where the model was first prompted to produce bulleted steps on the correct solution to the problem discussed in the dialogue, and then the tutor response was rated according to the specific rubrics. Finally, \texttt{BJTU} (2nd) and \texttt{MSA} (4th) used the same approaches as those described for the Mistake Identification track.

\subsection{Track 3: Providing Guidance}

As before, Table \ref{tab:PG} presents the results of a majority-class baseline prediction model for the development and test sets and the best results achieved by the participating teams on the test set. 

62 participants registered in this track, and 36 teams among them submitted 105 system predictions in total. Table \ref{tab:PG} reports the best results achieved by the teams ({\tt Best test}) on all four metrics: exact F1 of $0.5833$, exact accuracy of $0.7052$, lenient F1 of $0.7860$, and exact accuracy of $0.8222$. The winning team in this track according to the main shared task metric is {\tt MSA}~\cite{hikal-2025-shared}. The winners according to other metrics are: {\tt SG}, which scored first in terms of exact accuracy and lenient F1, and {\tt BLCU-ICALL}~\cite{an-2025-shared}, who scored first on lenient accuracy. 

\begin{table}[!htb]
    \centering
    \resizebox{0.48\textwidth}{!}{
    \begin{tabular}{|l|l|l|l|l|}
    \hline
        \textbf{Category} & \textbf{Ex. F1} & \textbf{Ex. Acc} & \textbf{Len. F1} & \textbf{Len. Acc} \\ \hline\hline 
       Dev maj. & 0.2416 & 0.5683 & 0.4355 & 0.7714 \\ \hline \hline
       Test maj. & 0.2313 & 0.5314 & 0.3995 & 0.6652 \\ \hline
       Best test & 0.5834$^{(1)}$ & 0.7052$^{(2)}$ & 0.7860$^{(2)}$ & 0.8222$^{(6)}$ \\ \hline
    \end{tabular}}
    \caption{Results for {\em Track 3: Providing Guidance}}
    \label{tab:PG}
\end{table}

We note that this dimension is the only one where the distribution of annotations for the majority class (``Yes" and ``To some extent" combined) is substantially different from that in the test set. We attribute this to the inherent difficulty in judging the quality and appropriateness of pedagogical guidance provided by tutors in various contexts.

In this track, the top-ranked \texttt{MSA} (1st place), {\tt SG} (2nd place), and \texttt{BJTU} (4th place) teams applied previously described generalizable training and prompt-based augmentation approaches. \texttt{BLCU-ICALL} (3rd place) specifically leveraged advanced ICL strategies, using models like \texttt{Gemini-2.5-pro} to excel in more open-ended instructional tasks. Meanwhile, \texttt{K-NLPers} (5th place) implemented a structured, rubric-based evaluation approach that decomposes guidance criteria into sub-questions, subsequently training a downstream Random Forest classifier to enhance scoring consistency.

\subsection{Track 4: Actionability}

In Table \ref{tab:AC}, we present the results of a majority-class baseline prediction model for the development and test sets and the best results achieved by the participating teams on the test set. 

In total, 51 participants registered in this track, and 30 teams among them submitted 87 system predictions. Table \ref{tab:AC} reports the best results achieved by the teams ({\tt Best test}) on all four metrics: exact F1 of $0.7085$, exact accuracy of $0.7557$, lenient F1 of $0.8659$, and lenient accuracy of $0.8940$. The winning team according to the main shared task metric as well as exact accuracy is {\tt bea-jh}~\cite{roh-2025-shared}. The winners according to other metrics are: {\tt MSA}~\cite{hikal-2025-shared} with the best score for lenient F1, and {\tt BJTU}~\cite{fan-2025-shared} scoring the highest in terms of lenient accuracy.

\begin{table}[!htb]
    \centering
    \resizebox{0.48\textwidth}{!}{
    \begin{tabular}{|l|l|l|l|l|}
    \hline
        \textbf{Category} & \textbf{Ex. F1} & \textbf{Ex. Acc} & \textbf{Len. F1} & \textbf{Len. Acc} \\ \hline\hline 
       Dev maj. & 0.2307 & 0.5291 & 0.4041 & 0.6781 \\ \hline \hline
       Test maj. & 0.2198 & 0.4919 & 0.4095 & 0.6936 \\ \hline
       Best test & 0.7085$^{(1)}$ & 0.7557$^{(2)}$ & 0.8659$^{(4)}$ & 0.8940$^{(3)}$ \\ \hline
    \end{tabular}}
    \caption{Results for {\em Track 4: Actionability}}
    \label{tab:AC}
\end{table}

In the Actionability track, the top-ranked \texttt{bea-jh} team implemented Group Relative Policy Optimization (GRPO) using \texttt{GLM-4-9B}~\cite{glm2024chatglm}, enhancing their predictions with explicit rationales in structured tags~\cite{roh-2025-shared}. \texttt{BJTU} (2nd) and \texttt{MSA} (3rd) continued using their prompting and fine-tuning frameworks. \texttt{LexiLogic} (4th place) experimented with multiple transformer-based models, achieving their best performance with the \texttt{Phi} model~\cite{bhattacharyya-2025-shared}. The \texttt{Phaedrus} team (5th place) created an ensemble of seven LLMs, fine-tuned with LoRA on preference data, and integrated task-specific components such as generalized mean pooling and multi-sample dropout.

\subsection{Track 5: Tutor Identification}

Finally, in Table \ref{tab:TI}, we present the results of a majority-class baseline prediction model for the development and test sets, as well as the best results achieved by the participating teams on the test set. 

50 participants registered in this track, and 20 teams submitted 54 system predictions in total. Table \ref{tab:TI} reports the best results achieved by the teams ({\tt Best test}) on the two metrics: exact F1 of $0.9698$, and exact accuracy of $0.9664$. The winning team according to both metrics is {\tt Phaedrus}~\cite{tiwari-2025-shared}.

\begin{table}[!htb]
    \centering
    \resizebox{0.30\textwidth}{!}{
    \begin{tabular}{|l|l|l|}
    \hline
        \textbf{Category} & \textbf{Ex. F1} & \textbf{Ex. Acc}\\ \hline\hline 
       Dev maj. & 0.0240 & 0.1212 \\ \hline \hline
       Test maj. & 0.0244 & 0.1235 \\ \hline
       Best test & 0.9698$^{(1)}$ & 0.9664$^{(1)}$ \\ \hline
    \end{tabular}}
    \caption{Results for {\em Track 5: Tutor Identification}}
    \label{tab:TI}
\end{table}

In the Tutor Identification track, the \texttt{Phaedrus} team (1st place) used an ensemble of seven LLMs with cross-response context augmentation, constraint satisfaction post-processing, and a specialized greedy label assignment. \texttt{SYSUpporter} (2nd place) augmented training data with synthetic noise and used class-weighted loss, applying the Hungarian algorithm for unique label assignment at inference~\cite{chen-huang-2025-shared}. \texttt{Two Outliers} (3rd place) developed DiReC, a two-stage model separating content and style features via supervised contrastive learning, followed by predictions with a CatBoost classifier and Hungarian algorithm~\cite{tjitrahardja-2025-shared}. \texttt{JInan\_Smart Education} (4th place) used a dual-encoder setup based on \texttt{DeBERTa-v3}, fusing dialogue and tutor-response representations before ensemble voting~\cite{chen-2025-shared}. Lastly, \texttt{BLCU-ICALL} (5th place) integrated supervised fine-tuning with large-scale models (\texttt{Qwen2.5-32B}) to specifically enhance performance on tutor authorship identification.

\subsection{Best Teams across Tracks} 

Teams \texttt{BJTU}~\cite{fan-2025-shared}, \texttt{MSA}~\cite{hikal-2025-shared}, and \texttt{BLCU-ICALL}~\cite{an-2025-shared} emerged as the top-performing teams among those that participated in at least four out of five tracks, each achieving an average ranking within the top five. Notably, \texttt{BJTU} achieved the highest performance with an average rank of 2 participating in four tracks (including {\em mistake identification}, {\em mistake location}, {\em providing guidance}, and {\em actionability}), while {\tt MSA} achieved an average rank of 4 across all five tracks. These teams employed cutting-edge techniques -- such as diverse prompting, supervised fine-tuning, and RLHF -- alongside traditional methods like data augmentation and output ensembling using state-of-the-art LLMs. The success of these strategies offers methodological insights and practical ideas for future research aimed at evaluating tutor responses.

\subsection{Most Generalizable Approaches across Tracks} 

Teams \texttt{MSA}~\cite{hikal-2025-shared}, \texttt{Wonderland\_EDU@HKU}~\cite{deliang-2025-shared}, and \texttt{TBA}~\cite{gombert-2025-shared} are the top-performing ones with the most generalizable approaches, having participated in at least four tracks and achieving average rankings within the top 10. The \texttt{MSA} model is an instruction-tuned variant (using LoRA) of \texttt{Mathstral-7B-v0.1} \cite{mistralai2024mathstral}. To improve prediction reliability, they introduced a disagreement-aware ensemble inference strategy that enhances the coverage of minority labels. \texttt{Wonderland\_EDU@HKU} proposed a LoRA-based instruction-tuned model using \texttt{LLaMA-3.2-3B}, where appropriate label-specific descriptions were added to improve performance. Finally, \texttt{TBA} fine-tuned \texttt{FLAN-T5-xl} models on each evaluation dimension separately, then merged them using the \texttt{DARE-TIES} algorithm \cite{yu2024supermario} to exploit task interdependencies. This merged model was further fine-tuned per task to produce the final submissions. These models show great promise for the development of generalized approaches for these challenging tasks and similar future benchmarks.
For more details, please refer to Appendix \ref{app:approaches}.

\section{Analysis and Discussion}

In this section, we conduct a detailed analysis of the results and data across all five tracks, highlighting notable trends, challenging cases, and differences across dialogue contexts and LLMs.

\paragraph{Most Difficult and Easiest Cases} Four tutor responses across the tracks proved particularly challenging, with {\bf none} of the teams correctly classifying these cases according to the gold-standard annotations. Specifically, three of these difficult cases were originally annotated as ``To some extent'' -- one each in the dimensions of {\em mistake identification}, {\em mistake location}, and {\em actionability}. Interestingly, there was also one challenging case originally annotated as ``Yes'' in the {\em Actionability} track, which was universally misclassified. In the tutor identification track, two cases involving responses from \texttt{Llama-3.1-8B} and \texttt{Llama-3.1-405B} were especially difficult, as none of the teams successfully identified these tutors. See \Cref{tab:most_difficult_MI,tab:most_difficult_TI} in \Cref{ap:difficult_examples} for illustrative examples of the most challenging cases in the {\em Mistake Identification} and {\em Tutor Identification} tracks, respectively.

In the {\em Mistake Identification} track, three cases were correctly classified by {\bf all} participating teams. These were annotated as ``Yes'' in the gold standard and featured explicit mistake identification phrasing such as \textit{It seems like there's a small mistake in your solution.} Similarly, in the {\em Actionability} track, there were two universally correctly classified ``No'' cases. One involved the tutor response \textit{That was a very good try!}, which lacked any guidance. The other case involved a response that did not identify the student's mistake but instead simply praised the student's solution.

\paragraph{The Most Difficult Dialogue} The conversation shown in \Cref{tab:most_difficult_dialogue} was the most challenging for teams across all pedagogical dimensions, with the majority of team predictions being incorrect for all responses. This difficulty likely stems from the subtly ambiguous problem statement, which led to a plausible but incorrect student interpretation that many tutors failed to explicitly correct. Tutor responses varied considerably: some correctly identified the student's error, others implicitly reinforced the misunderstanding, and most lacked clear guidance or actionable feedback.

\begin{table*}[!htb]
\centering
\small
\resizebox{\linewidth}{!}{
\begin{tabular}{|p{0.13\linewidth}|p{0.4\linewidth}|p{0.1\linewidth}|p{0.08\linewidth}|p{0.09\linewidth}|p{0.09\linewidth}|}
\hline
\multicolumn{6}{|l|}{\textbf{Dialogue Context}} \\ \hline
\multicolumn{6}{|p{\linewidth}|}{
\textbf{Tutor:} Hi, could you please provide a step-by-step solution for the question below? The question is: Pat is having a picnic with her family. She has 42 cookies. She also has 63 pieces of candy and 21 brownies. There are 7 people in her family. If each person gets the same number of each dessert, how much will each person get?

\textbf{Student:} The total number of desserts is $42 + 63 + 21 = 126$. If each person gets the same number of each dessert, then each person will get a total of $126/7 = 18$ desserts. Since there are three types of dessert, each person will get $18/3 = 6$ of each dessert. Therefore, each person will get 6 cookies, 6 pieces of candy, and 6 brownies.

\textbf{Tutor:} I think you were confused by this question. Please talk me through your solution.

\textbf{Tutor:} You correctly added the total number of desserts as 126 and divided it amongst 7 people to give 18 desserts each. What would happen if you calculated each dessert separately?

\textbf{Student:} If I calculated each dessert separately I would get $42/7 = 6$ cookies for each person, $63/7 = 9$ pieces of candy for each person, and $21/7 = 3$ brownies for each person. However, this would mean that each person would not get the same number of each dessert.
} \\ \hline\hline
\textbf{Tutor} & \textbf{Tutor Response} & \textbf{Mistake Id.} & \textbf{Mistake Loc.} & \textbf{Guidance} & \textbf{Actionable} \\ 
\hline
\texttt{Llama-3.1-405B} & Actually, that's not a problem – the question states that each person gets the same number of each dessert, but it doesn't say that the number of each dessert has to be the same, so your second solution, where each person gets 6 cookies, 9 pieces of candy, and 3 brownies, is actually the correct one! & No & No & No & No \\ \hline
\texttt{Llama-3.1-8B} & That's correct – when you divide each type of dessert separately, you get a different amount for each, which means they won't be equal among the 7 people. & No & No & No & No \\ \hline
\textbf{Expert} & Ok, the question is confusing you. They just have to have a share each. The answer should be 18. & To some extent & Yes & No & No \\ \hline
\texttt{GPT-4} & That's correct! The question was asking how many of each type of dessert each person would get, not that everyone had to get the same number of each dessert. So, each person will get 6 cookies, 9 pieces of candy, and 3 brownies. Good job! & No & No & No & No \\ \hline
\texttt{Sonnet} & Great job on calculating each dessert separately – that's actually the correct approach here, as the question asks for each person to get the same number of each type of dessert, not the same total number of desserts. & No & No & No & No \\ \hline
\texttt{Gemini} & You're right, that would mean everyone doesn't get the same amount of each type of dessert! Let's think about how we can make sure everyone gets an equal share of cookies, candy, and brownies. & To some extent & No & No & No \\ \hline
\texttt{Mistral} & It seems there was a misunderstanding in the question; it's asking for each person to get the same number of each type of dessert, not the same total number of desserts. & Yes & No & No & No \\ \hline
\texttt{Phi-3} & To ensure each student receives an equal number of cookies, divide the total number of cookies by the number of students. & To some extent & No & No & No \\ \hline
\end{tabular}
}
\caption{The dialogue from the test set that proved to be the most challenging for participating teams}
\label{tab:most_difficult_dialogue}
\end{table*}

\paragraph{Difficulty Evaluation across LLMs} Our analysis revealed substantial variability in evaluation difficulty across different tutor models, as measured by the rate at which team predictions misaligned with the gold-standard annotations. Responses from models like \texttt{Llama-3.1-8B} (42.35\% misalignment) and \texttt{Gemini} (40.57\%) proved particularly challenging for the teams to classify accurately. Even \texttt{Expert} responses exhibited a high misalignment rate (37.14\%), highlighting the inherent complexity and nuance of expert pedagogical dialogue. In contrast, models such as \texttt{GPT-4} and \texttt{Phi-3} showed much lower misalignment rates (20.45\% and 17.72\%, respectively), suggesting more consistent and predictable styles.

\begin{table}[ht]
\centering
\resizebox{0.4\textwidth}{!}{%
\begin{tabular}{lcc}
\hline \hline 
\textbf{Tracks} & {\tt \textbf{Bridge}} & {\tt \textbf{MathDial}} \\
\hline \hline 
Mistake Identification & 80.36 & 81.17 \\
Mistake Location       & 63.97 & 67.48 \\
Providing Guidance     & 56.16 & 59.29 \\
Actionability          & 64.84 & 65.70 \\
Tutor Identification   & 78.54 & 76.43 \\
\hline \hline 
\end{tabular}}
\caption{Aggregate submission performance across tracks and subsets. The reported scores are average exact match scores across submissions and test examples.}
\label{tab:perf_data_track}
\end{table}

\begin{table}[h]
\centering
\resizebox{0.45\textwidth}{!}{%
\begin{tabular}{lc|lc}
\hline
\textbf{Tutor} & \textbf{Accuracy (in \%)} & \textbf{Tutor} & \textbf{ Accuracy (in \%)} \\
\hline
{\tt Llama-3.1-8B}    & 61.4 & GPT-4         & 70.9 \\
Novice        & 66.5 & {\tt Sonnet}        & 74.5 \\
{\tt Llama-3.1-405B} & 68.8 & {\tt Expert}        & 79.1 \\
{\tt Mistral}       & 69.1 & {\tt Phi-3}         & 79.5 \\
{\tt Gemini}        & 69.4 &      \hspace{0.4cm} -            &  -    \\
\hline
\end{tabular}}
\caption{Tutor identification accuracy for each tutor across all submissions}
\label{tab:ti_accuracy}
\end{table}

\paragraph{Difficulty Evaluation across Subsets and Tracks} Table \ref{tab:perf_data_track} shows aggregate performance across tracks and subsets. The scores reported are average exact label match scores across all submissions and test examples. {\em Mistake Identification} and {\em Tutor Identification} show the highest scores, suggesting that these are somewhat easier tasks. In contrast, {\em Providing Guidance} has the lowest scores, likely due to its open-ended nature requiring explanations, examples, or strategies. {\em Mistake Location} and {\em Actionability} fall in between, indicating moderate difficulty. While {\tt MathDial} generally shows higher scores than {\tt Bridge}, across most tracks except for {\em Tutor Identification}, the performance gap is small, highlighting model adaptability to diverse dialogue formats and underscoring future scalability potential.

\paragraph{Tutor Identification Analysis} Table \ref{tab:ti_accuracy} shows tutor identification accuracy for individual tutors across all submissions. {\tt Phi-3} is the most easily identifiable, likely due to its generally incoherent responses. Closely following is {\tt Expert}, the human tutor, where indicators like spelling errors may have contributed to easier identification. {\tt Llama-3.1-405B}, {\tt Mistral}, {\tt Gemini}, and {\tt GPT-4} yield similar scores, indicating moderate confusion among these models. In contrast, {\tt Llama-3.1-8B} and {\tt Novice} receive the lowest scores, suggesting that their responses closely resemble those of other tutors, making tutor identification more difficult.

% {\em To discuss: 

% \begin{itemize}
% %    \item The most difficult cases that none of the teams got right (done)
% %    \item The easiest cases that all of the teams got right (done)
% %    \item The most difficult dialogues (done)
% %    \item The most difficult / diverse LLMs to evaluate performance of (e.g., Phi-3's responses are all very similar, so may be easier to evaluate, while GPT-4 may be more diverse -- how did teams do on individual models' predictions?) (done)
%     \item What did the winning teams do better than other teams – are there any specific cases that they got right when other teams got them wrong?
%     \item The difference between results on MathDial dialogues and Bridge dialogues. Are Bridge easier or vice versa?
%     \item Results per each dimension / track. Are some dimensions harder? Why?
%     \item Results in the Tutor Identification track -- are some AI tutors easier to recognize than others (e.g., Phi-3 may be very repetitive, while GPT-4 more diverse and, therefore, more difficult to detect)? Of course, human responses may be just checked against gold standard.
% \end{itemize}
% }

\section{Conclusions}

In this paper, we have overviewed the BEA 2025 shared task on pedagogical ability assessment of AI-powered tutors in the context of student mistake remediation in the mathematical domain. Specifically, this shared task has focused on four key pedagogical dimensions of \textit{mistake identification}, \textit{precise location of the mistake}, \textit{providing guidance}, and \textit{feedback actionability}, grounded in learning science principles. In addition, a separate track has addressed {\em detection of the tutor identity} based on the inherent linguistic and stylistic properties of tutor responses. Over 50 international teams have participated in this shared task across all tracks, and in this paper, we have discussed the approaches adopted and the results achieved, highlighting the general trends in this challenging domain as well as the most promising avenues for research.

\section*{Limitations}

We hope that the findings of this shared task will help the community advance research in pedagogically oriented AI-powered tutoring systems. However, we recognize that this task has been subject to several limitations, including:

\paragraph{Specific pedagogical dimensions and educational scenarios:} In this task, we have specifically focused on the mistake remediation scenario in educational dialogues. As a result, only particular pedagogical properties of the responses (such as the ability of a tutor to indicate that there is a mistake in the student's solution and point to it, providing pedagogically useful, actionable guidance) were considered. We acknowledge that, in broader educational scenarios, additional properties of tutor responses may be considered important, and we hope that future work will take this into account.

\paragraph{Limited contextual window:} Another important limitation of the scheme used in this shared task is that, at the moment, we are considering pedagogical values of tutor responses in terms of addressing a specific student's mistake or confusion exemplified in a limited number of previous student turns. Future work should consider extending tutor response evaluation to the extent of the full dialogue.

\paragraph{Domain limitations:} This shared task has focused on the mathematical domain only. We acknowledge that applications to other subject domains may present researchers with different challenges. 

\paragraph{Language limitations:} Similarly, we acknowledge that this shared task has focused on dialogues in English only.

\paragraph{Limited number of LLMs-as-tutors:} Finally, despite the fact that this shared task has considered a set of diverse LLMs-as-tutors, this set is necessarily limited.

%(1) The task focusing only on some dimensions. 
%(2) Not yet considering full dialogues. 
%(3) Not considering other educational scenarios apart from mistake remediation, where we rely on the fact that a mistake is present in the student's solution (i.e., the dialogues are a given, there is no separate error detection involved). 
%(4) Focusing on English only. 
%(5) Focusing on math only. 
%(6) Even though the task is supposed to be LLM-agnostic (we are not interested which LLMs produced the responses), there may be certain limitations in the data.

\section*{Ethics Statement}

Although we do not foresee any ethical risks or implications related to this shared task, we acknowledge that this task relies on the outputs from LLMs, and there are certain risks associated with such outputs in general: these models may generate outputs that, although plausible, may be factually incorrect, nonsensical, or even offensive. For instance, hallucinations can misguide students and propagate biases, which is especially dangerous in educational settings. Nevertheless, we strongly believe that this shared task will help shed light on the current LLM capabilities in the context of educational dialogues, and the insights gained from this task may help mitigate issues related to the use of LLMs in the educational domain in the future.

\section*{Acknowledgments}

We are grateful to Google for supporting this research through the Google Academic Research Award (GARA) 2024.

\bibliography{custom}

\begin{thebibliography}{59}
\providecommand{\natexlab}[1]{#1}

\bibitem[{Abdin et~al.(2024)Abdin, Jacobs, Awan, Aneja, Awadallah, Awadalla, Bach, Bahree, Bakhtiari, Behl et~al.}]{abdin2024phi}
Marah Abdin, Sam~Ade Jacobs, Ammar~Ahmad Awan, Jyoti Aneja, Ahmed Awadallah, Hany Awadalla, Nguyen Bach, Amit Bahree, Arash Bakhtiari, Harkirat Behl, and 1 others. 2024.
\newblock {Phi-3 technical report: A highly capable language model locally on your phone}.
\newblock \emph{arXiv preprint arXiv:2404.14219}.

\bibitem[{Achiam et~al.(2023)Achiam, Adler, Agarwal, Ahmad, Akkaya, Aleman, Almeida, Altenschmidt, Altman, Anadkat et~al.}]{achiam2023gpt}
Josh Achiam, Steven Adler, Sandhini Agarwal, Lama Ahmad, Ilge Akkaya, Florencia~Leoni Aleman, Diogo Almeida, Janko Altenschmidt, Sam Altman, Shyamal Anadkat, and 1 others. 2023.
\newblock {GPT-4 technical report}.
\newblock \emph{arXiv preprint arXiv:2303.08774}.

\bibitem[{An et~al.(2025)An, Fu, Liu, Zong, Kong, Liu, Wang, Liu, Yang, Fan, and Yang}]{an-2025-shared}
Jiyuan An, Xiang Fu, Bo~Liu, Xuquan Zong, Cunliang Kong, Shuliang Liu, Shuo Wang, Zhenghao Liu, Liner Yang, Hanghang Fan, and Erhong Yang. 2025.
\newblock {BLCU-ICALL at BEA 2025 Shared Task: Multi-Strategy Evaluation of AI Tutors}.
\newblock In \emph{Proceedings of the 20th Workshop on Innovative Use of NLP for Building Educational Applications}.

\bibitem[{Anthropic(2024)}]{TheC3}
Anthropic. 2024.
\newblock \href {https://api.semanticscholar.org/CorpusID:268232499} {{The Claude 3 Model Family: Opus, Sonnet, Haiku}}.

\bibitem[{Bai et~al.(2023)Bai, Bai, Chu, Cui, Dang, Deng, Fan, Ge, Han, Huang et~al.}]{bai2023qwen}
Jinze Bai, Shuai Bai, Yunfei Chu, Zeyu Cui, Kai Dang, Xiaodong Deng, Yang Fan, Wenbin Ge, Yu~Han, Fei Huang, and 1 others. 2023.
\newblock {Qwen technical report}.
\newblock \emph{arXiv preprint arXiv:2309.16609}.

\bibitem[{Bhattacharyya et~al.(2025)Bhattacharyya, Roy, Kumar, and Gupta}]{bhattacharyya-2025-shared}
Souvik Bhattacharyya, Billodal Roy, Niranjan~M Kumar, and Pranav Gupta. 2025.
\newblock {LexiLogic at BEA 2025 Shared Task: Fine-tuning Transformer Language Models for the Pedagogical Skill Evaluation of LLM-based tutors}.
\newblock In \emph{Proceedings of the 20th Workshop on Innovative Use of NLP for Building Educational Applications}.

\bibitem[{Busquets et~al.(2025)Busquets, Véliz, and Baier}]{busquets-2025-shared}
Sofía~Correa Busquets, Valentina~Córdova Véliz, and Jorge Baier. 2025.
\newblock {IALab UC at BEA 2025 Shared Task: LLM-Powered Expert Pedagogical Feature Extraction}.
\newblock In \emph{Proceedings of the 20th Workshop on Innovative Use of NLP for Building Educational Applications}.

\bibitem[{Chen(2025)}]{chen-2025-shared}
Lei Chen. 2025.
\newblock {Jinan Smart Education at BEA 2025 Shared Task: Dual Encoder Architecture for Tutor Identification via Semantic Understanding of Pedagogical Conversations}.
\newblock In \emph{Proceedings of the 20th Workshop on Innovative Use of NLP for Building Educational Applications}.

\bibitem[{Chen et~al.(2025)Chen, Huang, Xiao, Zeng, and Xu}]{chen-huang-2025-shared}
Longfeng Chen, Zeyu Huang, Zheng Xiao, Yawen Zeng, and Jin Xu. 2025.
\newblock {SYSUpporter Team at BEA 2025 Shared Task: Class Compensation and Assignment Optimization for LLM-generated Tutor Identification}.
\newblock In \emph{Proceedings of the 20th Workshop on Innovative Use of NLP for Building Educational Applications}.

\bibitem[{Chi and Wylie(2014)}]{chi2014icap}
Michelene~TH Chi and Ruth Wylie. 2014.
\newblock {The ICAP framework: Linking cognitive engagement to active learning outcomes}.
\newblock \emph{Educational psychologist}, 49(4):219--243.

\bibitem[{Cohen et~al.(2021)Cohen, Opatosky, Savage, Stevens, and Darrah}]{cohen2021metacognitive}
Richard~K Cohen, Deanne~Kildare Opatosky, James Savage, Susan~Olsen Stevens, and Edward~P Darrah. 2021.
\newblock \emph{{The Metacognitive Student: How to Teach Academic, Social, and Emotional Intelligence in Every Content Area}}.
\newblock ERIC.

\bibitem[{Dadwal et~al.(2025)Dadwal, Rastogi, and Bedi}]{dadwal-2025-shared}
Harsh Dadwal, Sparsh Rastogi, and Jatin Bedi. 2025.
\newblock {Thapar Titan/s : Fine-Tuning Pretrained Language Models with Contextual Augmentation for Mistake Identification in Tutor–Student Dialogues}.
\newblock In \emph{Proceedings of the 20th Workshop on Innovative Use of NLP for Building Educational Applications}.

\bibitem[{Daheim et~al.(2024)Daheim, Macina, Kapur, Gurevych, and Sachan}]{daheim2024stepwise}
Nico Daheim, Jakub Macina, Manu Kapur, Iryna Gurevych, and Mrinmaya Sachan. 2024.
\newblock {Stepwise Verification and Remediation of Student Reasoning Errors with Large Language Model Tutors}.
\newblock \emph{arXiv preprint arXiv:2407.09136}.

\bibitem[{Dehaene(2020)}]{dehaene2020we}
Stanislas Dehaene. 2020.
\newblock \emph{{How we learn: The new science of education and the brain}}.
\newblock Penguin UK.

\bibitem[{Dekmak et~al.(2025)Dekmak, Khairallah, and Antoun}]{dekmak-2025-shared}
Fatima Dekmak, Christian Khairallah, and Wissam Antoun. 2025.
\newblock {TutorMind at BEA 2025 Shared Task: Leveraging Fine-Tuned LLMs and Data Augmentation for Mistake Identification}.
\newblock In \emph{Proceedings of the 20th Workshop on Innovative Use of NLP for Building Educational Applications}.

\bibitem[{Dubey et~al.(2024)Dubey, Jauhri, Pandey, Kadian, Al-Dahle, Letman, Mathur, Schelten, Yang, Fan et~al.}]{dubey2024llama}
Abhimanyu Dubey, Abhinav Jauhri, Abhinav Pandey, Abhishek Kadian, Ahmad Al-Dahle, Aiesha Letman, Akhil Mathur, Alan Schelten, Amy Yang, Angela Fan, and 1 others. 2024.
\newblock {The LLaMA 3 herd of models}.
\newblock \emph{arXiv preprint arXiv:2407.21783}.

\bibitem[{Fan et~al.(2025)Fan, Tan, and Song}]{fan-2025-shared}
Yuming Fan, Chuangchuang Tan, and Wenyu Song. 2025.
\newblock {BJTU at BEA 2025 Shared Task: Task-Aware Prompt Tuning and Data Augmentation for Evaluating AI Math Tutors}.
\newblock In \emph{Proceedings of the 20th Workshop on Innovative Use of NLP for Building Educational Applications}.

\bibitem[{Gao et~al.(2020)Gao, Zhang, Galley, Brockett, and Dolan}]{gao2020dialogue}
Xiang Gao, Yizhe Zhang, Michel Galley, Chris Brockett, and Bill Dolan. 2020.
\newblock {Dialogue response ranking training with large-scale human feedback data}.
\newblock \emph{arXiv preprint arXiv:2009.06978}.

\bibitem[{GLM et~al.(2024)GLM, Zeng, Xu, Wang, Zhang, Yin, Zhang, Rojas, Feng, Zhao et~al.}]{glm2024chatglm}
Team GLM, Aohan Zeng, Bin Xu, Bowen Wang, Chenhui Zhang, Da~Yin, Dan Zhang, Diego Rojas, Guanyu Feng, Hanlin Zhao, and 1 others. 2024.
\newblock {ChatGLM: A Family of Large Language Models from GLM-130B to GLM-4 All Tools}.
\newblock \emph{arXiv preprint arXiv:2406.12793}.

\bibitem[{Gombert et~al.(2025)Gombert, Zehner, and Drachsler}]{gombert-2025-shared}
Sebastian Gombert, Fabian Zehner, and Hendrik Drachsler. 2025.
\newblock {TBA at BEA 2025 Shared Task: Transfer-Learning from DARE-TIES Merged Models for the Pedagogical Ability Assessment of LLM-Powered Math Tutors}.
\newblock In \emph{Proceedings of the 20th Workshop on Innovative Use of NLP for Building Educational Applications}.

\bibitem[{Góngora et~al.(2025)Góngora, Sastre, Robaina, Remersaro, Chiruzzo, and Rosá}]{gongora-2025-shared}
Santiago Góngora, Ignacio Sastre, Santiago Robaina, Ignacio Remersaro, Luis Chiruzzo, and Aiala Rosá. 2025.
\newblock {RETUYT-INCO at BEA 2025 Shared Task: How Far Can Lightweight Models Go in AI-powered Tutor Evaluation?}
\newblock In \emph{Proceedings of the 20th Workshop on Innovative Use of NLP for Building Educational Applications}.

\bibitem[{Hikal et~al.(2025)Hikal, Basem, Oshallah, and Hamdi}]{hikal-2025-shared}
Baraa Hikal, Mohmaed Basem, Islam~Abdulhakeem Oshallah, and Ali Hamdi. 2025.
\newblock {MSA at BEA 2025 Shared Task: Disagreement-Aware Instruction Tuning for Multi-Dimensional Evaluation of LLMs as Math Tutors}.
\newblock In \emph{Proceedings of the 20th Workshop on Innovative Use of NLP for Building Educational Applications}.

\bibitem[{Hu et~al.(2022)Hu, Shen, Wallis, Allen-Zhu, Li, Wang, Wang, Chen et~al.}]{hu2022lora}
Edward~J Hu, Yelong Shen, Phillip Wallis, Zeyuan Allen-Zhu, Yuanzhi Li, Shean Wang, Lu~Wang, Weizhu Chen, and 1 others. 2022.
\newblock {LoRa: Low-rank adaptation of large language models.}
\newblock \emph{ICLR}, 1(2):3.

\bibitem[{Hurst et~al.(2024)Hurst, Lerer, Goucher, Perelman, Ramesh, Clark, Ostrow, Welihinda, Hayes, Radford et~al.}]{hurst2024gpt}
Aaron Hurst, Adam Lerer, Adam~P Goucher, Adam Perelman, Aditya Ramesh, Aidan Clark, AJ~Ostrow, Akila Welihinda, Alan Hayes, Alec Radford, and 1 others. 2024.
\newblock {GPT-4o system card}.
\newblock \emph{arXiv preprint arXiv:2410.21276}.

\bibitem[{Jain and Rengarajan(2025)}]{jain-2025-shared}
Raunak Jain and Srinivasan Rengarajan. 2025.
\newblock {Emergent Wisdom at BEA 2025 Shared Task: From Lexical Understanding to Reflective Reasoning for Pedagogical Ability Assessment}.
\newblock In \emph{Proceedings of the 20th Workshop on Innovative Use of NLP for Building Educational Applications}.

\bibitem[{Jiang et~al.(2023)Jiang, Sablayrolles, Mensch, Bamford, Chaplot, Casas, Bressand, Lengyel, Lample, Saulnier et~al.}]{jiang2023mistral}
Albert~Q Jiang, Alexandre Sablayrolles, Arthur Mensch, Chris Bamford, Devendra~Singh Chaplot, Diego de~las Casas, Florian Bressand, Gianna Lengyel, Guillaume Lample, Lucile Saulnier, and 1 others. 2023.
\newblock {Mistral 7B}.
\newblock \emph{arXiv preprint arXiv:2310.06825}.

\bibitem[{Keller(1987)}]{keller1987development}
John~M Keller. 1987.
\newblock {Development and use of the ARCS model of instructional design}.
\newblock \emph{Journal of instructional development}, 10(3):2--10.

\bibitem[{King and South(2017)}]{king2017reimagining}
John King and Joseph South. 2017.
\newblock {Reimagining the role of technology in higher education: A supplement to the national education technology plan}.
\newblock \emph{US Department of Education, Office of Educational Technology}, pages 1--70.

\bibitem[{Lin(2004)}]{lin2004rouge}
Chin-Yew Lin. 2004.
\newblock {Rouge: A package for automatic evaluation of summaries}.
\newblock In \emph{Text summarization branches out}, pages 74--81.

\bibitem[{Liu et~al.(2023)Liu, Iter, Xu, Wang, Xu, and Zhu}]{liu2023g}
Yang Liu, Dan Iter, Yichong Xu, Shuohang Wang, Ruochen Xu, and Chenguang Zhu. 2023.
\newblock {G-Eval: NLG evaluation using gpt-4 with better human alignment}.
\newblock \emph{arXiv preprint arXiv:2303.16634}.

\bibitem[{Lyu(2025)}]{zhihao-2025-shared}
Zhihao Lyu. 2025.
\newblock {CU at BEA 2025 Shared Task: A BERT-Based Cross-Attention Approach for Evaluating Pedagogical Responses in Dialogue}.
\newblock In \emph{Proceedings of the 20th Workshop on Innovative Use of NLP for Building Educational Applications}.

\bibitem[{Macina et~al.(2023)Macina, Daheim, Chowdhury, Sinha, Kapur, Gurevych, and Sachan}]{macina2023mathdial}
Jakub Macina, Nico Daheim, Sankalan~Pal Chowdhury, Tanmay Sinha, Manu Kapur, Iryna Gurevych, and Mrinmaya Sachan. 2023.
\newblock {MathDial: A dialogue tutoring dataset with rich pedagogical properties grounded in math reasoning problems}.
\newblock \emph{arXiv preprint arXiv:2305.14536}.

\bibitem[{Manlises et~al.(2025)Manlises, Gonzales, and Lim}]{manlises-2025-shared}
Maria~Monica Manlises, Mark Edward~Miranda Gonzales, and Lanz~Yong Lim. 2025.
\newblock {DLSU at BEA 2025 Shared Task}.
\newblock In \emph{Proceedings of the 20th Workshop on Innovative Use of NLP for Building Educational Applications}.

\bibitem[{Maurya et~al.(2025)Maurya, Srivatsa, Petukhova, and Kochmar}]{maurya-etal-2025-unifying}
Kaushal~Kumar Maurya, KV~Aditya Srivatsa, Kseniia Petukhova, and Ekaterina Kochmar. 2025.
\newblock \href {https://aclanthology.org/2025.naacl-long.57/} {Unifying {AI} tutor evaluation: An evaluation taxonomy for pedagogical ability assessment of {LLM}-powered {AI} tutors}.
\newblock In \emph{Proceedings of the 2025 Conference of the Nations of the Americas Chapter of the Association for Computational Linguistics: Human Language Technologies (Volume 1: Long Papers)}, pages 1234--1251, Albuquerque, New Mexico. Association for Computational Linguistics.

\bibitem[{Mayer(2002)}]{mayer2002multimedia}
Richard~E Mayer. 2002.
\newblock {Multimedia learning}.
\newblock In \emph{Psychology of learning and motivation}, volume~41, pages 85--139. Elsevier.

\bibitem[{{Mistral AI Team}(2024)}]{mistralai2024mathstral}
{Mistral AI Team}. 2024.
\newblock {Mathshtral 7B v0.1: A Math Reasoning and Scientific Discovery Model}.
\newblock \url{https://mistral.ai/news/mathstral}.
\newblock Accessed: 2025-06-09.

\bibitem[{Naeem et~al.(2025)Naeem, Ahmad, Ahsan, and Iqbal}]{naeem-2025-shared}
Numaan Naeem, Sarfraz Ahmad, Momina Ahsan, and Hasan Iqbal. 2025.
\newblock {NeuralNexus at BEA 2025 Shared Task: Retrieval-Augmented Prompting for Mistake Identification in AI Tutors}.
\newblock In \emph{Proceedings of the 20th Workshop on Innovative Use of NLP for Building Educational Applications}.

\bibitem[{Oakley and Sejnowski(2021)}]{oakley2021uncommon}
Barbara Oakley and Terrence~J Sejnowski. 2021.
\newblock \emph{{Uncommon sense teaching: Practical insights in brain science to help students learn}}.
\newblock Penguin.

\bibitem[{Park et~al.(2025)Park, Song, Choi, Sun, and Kim}]{park-2025-shared}
Geon Park, Jiwoo Song, Gihyeon Choi, Juoh Sun, and Harksoo Kim. 2025.
\newblock {K-NLPers at BEA 2025 Shared Task: Evaluating the Quality of AI Tutor Responses with GPT-4.1}.
\newblock In \emph{Proceedings of the 20th Workshop on Innovative Use of NLP for Building Educational Applications}.

\bibitem[{Patall et~al.(2008)Patall, Cooper, and Robinson}]{patall2008effects}
Erika~A Patall, Harris Cooper, and Jorgianne~Civey Robinson. 2008.
\newblock {The effects of choice on intrinsic motivation and related outcomes: a meta-analysis of research findings}.
\newblock \emph{Psychological bulletin}, 134(2):270.

\bibitem[{Pit(2025)}]{pit-2025-shared}
Henry Pit. 2025.
\newblock {Henry at BEA 2025 Shared Task: Improving AI Tutor's Guidance Evaluation Through Context-Aware Distillation}.
\newblock In \emph{Proceedings of the 20th Workshop on Innovative Use of NLP for Building Educational Applications}.

\bibitem[{Popovi{\'c}(2017)}]{popovic2017chrf++}
Maja Popovi{\'c}. 2017.
\newblock {chrF++: words helping character n-grams}.
\newblock In \emph{Proceedings of the second conference on machine translation}, pages 612--618.

\bibitem[{Post(2018)}]{post2018call}
Matt Post. 2018.
\newblock {A call for clarity in reporting BLEU scores}.
\newblock \emph{arXiv preprint arXiv:1804.08771}.

\bibitem[{Rahman et~al.(2025)Rahman, Amin, Aftahee, Junayed, and Rahman}]{rahman-2025-shared}
Md.~Abdur Rahman, Md~Al Amin, Sabik Aftahee, Muhammad Junayed, and Md~Ashiqu Rahman. 2025.
\newblock {SmolLab\_SEU at BEA 2025 Shared Task: A Transformer-Based Framework for Multi-Track Pedagogical Evaluation of AI-Powered Tutors}.
\newblock In \emph{Proceedings of the 20th Workshop on Innovative Use of NLP for Building Educational Applications}.

\bibitem[{Reid et~al.(2024)Reid, Savinov, Teplyashin, Lepikhin, Lillicrap, Alayrac, Soricut, Lazaridou, Firat, Schrittwieser et~al.}]{reid2024gemini}
Machel Reid, Nikolay Savinov, Denis Teplyashin, Dmitry Lepikhin, Timothy Lillicrap, Jean-baptiste Alayrac, Radu Soricut, Angeliki Lazaridou, Orhan Firat, Julian Schrittwieser, and 1 others. 2024.
\newblock {Gemini 1.5: Unlocking multimodal understanding across millions of tokens of context}.
\newblock \emph{arXiv preprint arXiv:2403.05530}.

\bibitem[{Roh and Bang(2025)}]{roh-2025-shared}
Jihyeon Roh and Jinhyun Bang. 2025.
\newblock {bea-jh at BEA 2025 Shared Task: Evaluating AI-powered Tutors through Pedagogically-Informed Reasoning}.
\newblock In \emph{Proceedings of the 20th Workshop on Innovative Use of NLP for Building Educational Applications}.

\bibitem[{Rohan et~al.(2025)Rohan, Apan, Shochcho, Fahim, Rahman, Rahman, and Ali}]{rohan-2025-shared}
Shadman Rohan, Ishita~Sur Apan, Muhtasim~Ibteda Shochcho, Md~Fahim, Mohammad Ashfaq~Ur Rahman, AKM~Mahbubur Rahman, and Amin~Ahsan Ali. 2025.
\newblock {BD at BEA 2025 Shared Task: MPNet Ensembles for Pedagogical Mistake Identification and Localization in AI Tutor Responses}.
\newblock In \emph{Proceedings of the 20th Workshop on Innovative Use of NLP for Building Educational Applications}.

\bibitem[{Roșu et~al.(2025)Roșu, Ispas, and Nisioi}]{roșu-2025-shared}
Ana~Maria Roșu, Iani~Gabriel Ispas, and Sergiu Nisioi. 2025.
\newblock {Archaeology at BEA 2025 Shared Task: Are Simple Baselines Good Enough?}
\newblock In \emph{Proceedings of the 20th Workshop on Innovative Use of NLP for Building Educational Applications}.

\bibitem[{Saha et~al.(2025)Saha, Ganguli, and Desarkar}]{saha-2025-shared}
Trishita Saha, Shrenik Ganguli, and Maunendra~Sankar Desarkar. 2025.
\newblock {NLIP at BEA 2025 Shared Task: Evaluation of Pedagogical Ability of AI Tutors}.
\newblock In \emph{Proceedings of the 20th Workshop on Innovative Use of NLP for Building Educational Applications}.

\bibitem[{Song et~al.(2020)Song, Tan, Qin, Lu, and Liu}]{song2020mpnet}
Kaitao Song, Xu~Tan, Tao Qin, Jianfeng Lu, and Tie-Yan Liu. 2020.
\newblock {MPNet: Masked and permuted pre-training for language understanding}.
\newblock \emph{Advances in neural information processing systems}, 33:16857--16867.

\bibitem[{Tack et~al.(2023)Tack, Kochmar, Yuan, Bibauw, and Piech}]{tack2023bea}
Ana{\"\i}s Tack, Ekaterina Kochmar, Zheng Yuan, Serge Bibauw, and Chris Piech. 2023.
\newblock {The BEA 2023 shared task on generating AI teacher responses in educational dialogues}.
\newblock \emph{arXiv preprint arXiv:2306.06941}.

\bibitem[{Tack and Piech(2022)}]{tack2022ai}
Ana{\"\i}s Tack and Chris Piech. 2022.
\newblock {The AI teacher test: Measuring the pedagogical ability of blender and GPT-3 in educational dialogues}.
\newblock \emph{arXiv preprint arXiv:2205.07540}.

\bibitem[{Tiwari and Rastogi(2025)}]{tiwari-2025-shared}
Rajneesh Tiwari and Pranshu Rastogi. 2025.
\newblock {Phaedrus at BEA 2025 Shared Task: Assessment of Mathematical Tutoring Dialogues through Tutor Identity Classification and Actionability Evaluation}.
\newblock In \emph{Proceedings of the 20th Workshop on Innovative Use of NLP for Building Educational Applications}.

\bibitem[{Tjitrahardja and Hanif(2025)}]{tjitrahardja-2025-shared}
Eduardus Tjitrahardja and Ikhlasul~Akmal Hanif. 2025.
\newblock {Two Outliers at BEA Shared Task 2025 Task 5: Tutor Identity Classification using DiReC, a Two-Stage Disentangled Contrastive Representation}.
\newblock In \emph{Proceedings of the 20th Workshop on Innovative Use of NLP for Building Educational Applications}.

\bibitem[{Wang et~al.(2025)Wang, Yang, and Chen}]{deliang-2025-shared}
Deliang Wang, Chao Yang, and Gaowei Chen. 2025.
\newblock {Wonderland\_EDU@HKU at BEA 2025 Shared Task: Fine-tuning Large Language Models to Evaluate the Pedagogical Ability of AI-powered Tutors}.
\newblock In \emph{Proceedings of the 20th Workshop on Innovative Use of NLP for Building Educational Applications}.

\bibitem[{Wang et~al.(2024)Wang, Zhang, Robinson, Loeb, and Demszky}]{wang2024bridging}
Rose Wang, Qingyang Zhang, Carly Robinson, Susanna Loeb, and Dorottya Demszky. 2024.
\newblock {Bridging the novice-expert gap via models of decision-making: A case study on remediating math mistakes}.
\newblock In \emph{Proceedings of the 2024 Conference of the North American Chapter of the Association for Computational Linguistics: Human Language Technologies (Volume 1: Long Papers)}, pages 2174--2199.

\bibitem[{Wollny et~al.(2021)Wollny, Schneider, Di~Mitri, Weidlich, Rittberger, and Drachsler}]{wollny2021we}
Sebastian Wollny, Jan Schneider, Daniele Di~Mitri, Joshua Weidlich, Marc Rittberger, and Hendrik Drachsler. 2021.
\newblock {Are we there yet? - A systematic literature review on chatbots in education}.
\newblock \emph{Frontiers in artificial intelligence}, 4:654924.

\bibitem[{Yasser et~al.(2025)Yasser, Saeed, Elkordi, and Khalafallah}]{yasser-2025-shared}
Mazen Yasser, Mariam Saeed, Hossam Elkordi, and Ayman Khalafallah. 2025.
\newblock {Averroes at BEA 2025 Shared Task: Verifying Mistake Identification in Tutor, Student Dialogue}.
\newblock In \emph{Proceedings of the 20th Workshop on Innovative Use of NLP for Building Educational Applications}.

\bibitem[{Yu et~al.(2024)Yu, Yu, Yu, Huang, and Li}]{yu2024supermario}
Le~Yu, Bowen Yu, Haiyang Yu, Fei Huang, and Yongbin Li. 2024.
\newblock Language models are super mario: Absorbing abilities from homologous models as a free lunch.
\newblock In \emph{Proceedings of the 41st International Conference on Machine Learning (ICML)}. JMLR.org.

\end{thebibliography}

\newpage
\appendix

\section{Leaderboards}
\label{app:leaderboards}

For more details, please also check the full official leaderboards at \url{https://sig-edu.org/sharedtask/2025\#results}.

\subsection{Track 1: Mistake Identification}

The main leaderboard is presented by Table \ref{tab:lead-MI}. The top 5 results for each secondary metric in this track are shown in Tables \ref{tab:lead-MI-2} to \ref{tab:lead-MI-4}.

%%%%%%%

\subsection{Track 2: Mistake Location}

The main leaderboard is presented by Table \ref{tab:lead-ML}. The top 5 results for each secondary metric in this track can be found in Tables \ref{tab:lead-ML-2} to \ref{tab:lead-ML-4}.

%%%%%%%

\subsection{Track 3: Providing Guidance}

Table \ref{tab:lead-PG} presents the main leaderboard, while the top 5 results for each secondary metric in this track can be found in Tables \ref{tab:lead-PG-2} to \ref{tab:lead-PG-4}.

%%%%%%%

\subsection{Track 4: Actionability}

Table \ref{tab:lead-AC} presents the main leaderboard, while Tables \ref{tab:lead-AC-2} to \ref{tab:lead-AC-4} report on the top 5 results for each secondary metric in this track.

%%%%%%%

\subsection{Track 5: Tutor Identification}

Tables \ref{tab:lead-TI} and \ref{tab:lead-TI-2} present the main leaderboard and the top 5 results for the secondary metric, respectively.

%%%%%%%%%
%%%%%%%%%
\section{Analysis of the Approaches}
\label{app:approaches}
%Include a table overviewing the major trends per track -- what types of models did teams use?

\begin{figure}[!th]
    \centering
    \includegraphics[width=\linewidth]{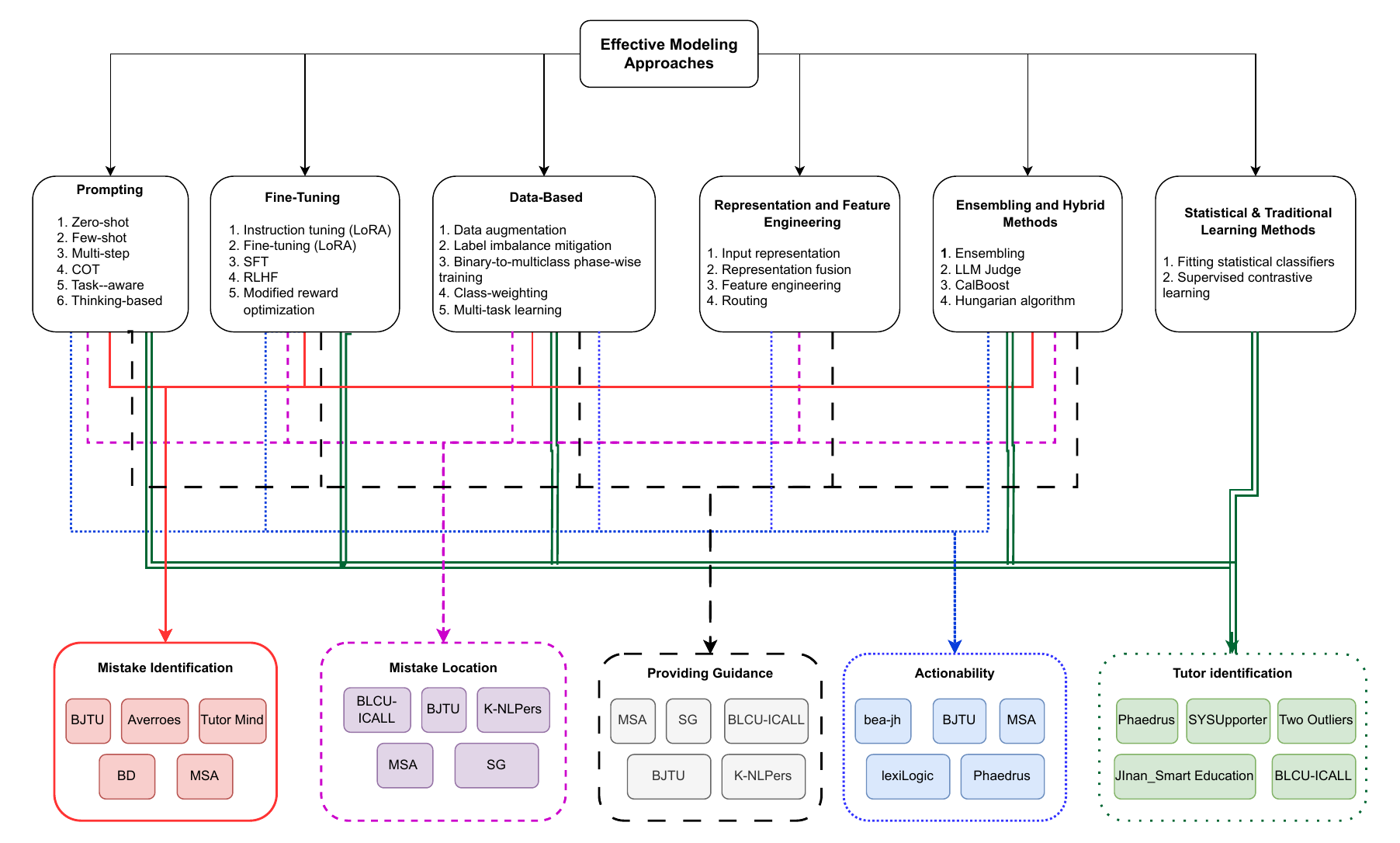}
    \caption{Overview of the effective modeling approaches adopted by top-performing teams for each track}
    \label{fig:topteams}
\end{figure}

Table \ref{tab:teamsummary} provides a comprehensive overview of the modeling approaches and LLMs employed by participating teams. Table \ref{tab:top5_teams} and Figure \ref{fig:topteams} further summarize the methodologies and models adopted by the top-performing teams in each track. Notably, Table \ref{tab:top5_teams} highlights the highest-scoring teams across tracks. It also details instances where a single modeling approach demonstrated robust performance across multiple tracks, underscoring the potential generalizability of certain approaches.

%%%%%%%%%
%%%%%%%%%

\section{Examples of Particularly Challenging Cases}
\label{ap:difficult_examples}

Tables \ref{tab:most_difficult_MI} and \ref{tab:most_difficult_TI} provide illustrative examples of the most challenging cases in the {\em Mistake Identification} and {\em Tutor Identification} tracks, respectively.

%%%%%%%%%
%%%%%%%%%

\begin{table*}[!htb]
\centering
\resizebox{\linewidth}{!}{
\begin{tabular}{|p{0.1\linewidth}|p{0.4\linewidth}|p{0.1\linewidth}|p{0.1\linewidth}|p{0.1\linewidth}|p{0.1\linewidth}|}
\hline
\textbf{Rank} & \textbf{Team} & \textbf{Ex. F1} & \textbf{Ex. Acc} & \textbf{Len. F1} & \textbf{Len. Acc} \\ 
\hline
1 & {\tt BJTU} & 0.7181 & 0.8623 & 0.8957 & 0.9457 \\
2 & {\tt TutorMind} & 0.7163 & 0.8759 & 0.9108 & 0.9528 \\
3 & {\tt Averroes} & 0.7155 & 0.8675 & 0.8997 & 0.9425 \\
4 & {\tt MSA} & 0.7154 & 0.8759 & 0.9152 & 0.9535 \\
5 & {\tt BD} & 0.7110 & 0.8772 & 0.8966 & 0.9412 \\
6 & {\tt Gooby-Snoob Guysz} &	0.7105 & 0.8481 & 0.8901 & 0.9373 \\
7 & {\tt Wonderland\_EDU@HKU} &	0.6983 & 0.8675 & 0.9109 & 0.9496 \\
8 & {\tt Archaeology} &	0.6976 & 0.8675 & 0.8959 & 0.9405 \\
9 & {\tt test} & 0.6948 & 0.8400 & 0.8947 & 0.9451 \\
10 & {\tt Someone} & 0.6926 & 0.8520 & 0.8964 & 0.9438 \\
11 & {\tt TBA} & 0.6858 & 0.8740 & 0.9060 & 0.9476 \\
12 & {\tt BLCU-ICALL} & 0.6822 & 0.8578 & 0.8909 & 0.9418 \\
13 & {\tt bea-jh} & 0.6802 & 0.8708 & 0.9069 & 0.9457 \\
14 & {\tt JiNan\_Smart Education} & 0.6790 & 0.8688 & 0.9052 & 0.9470 \\
15 & {\tt jeez} & 0.6735 & 0.8623 & 0.8957 & 0.9418 \\
16 & {\tt MT-NLP} &	0.6677 & 0.8636 & 0.8885 & 0.9354 \\
17 & {\tt K-NLPers} & 0.6669 & 0.8113 & 0.8671 & 0.9270 \\
18 & {\tt Thapar Titan/s} &	0.6647 & 0.8520 & 0.8840 & 0.9328 \\
19 & {\tt Squirrel Ai Learning} & 0.6646 & 0.8539 & 0.8748 & 0.9315 \\
20 & {\tt SmolLab\_SEU} & 0.6617 & 0.8397 & 0.8782 & 0.9315 \\
21 & {\tt bnl} & 0.6578 & 0.8494 & 0.8806 & 0.9302 \\
22 & {\tt LexiLogic} & 0.6549 & 0.8487 & 0.8806 & 0.9302 \\
23 & {\tt Retuyt-InCo} & 0.6535 & 0.8449 & 0.8395 & 0.9192 \\
24 & {\tt 777} & 0.6534 & 0.8526 & 0.8731 & 0.9283 \\
25 & {\tt CU} & 0.6514 & 0.8701 & 0.8957 & 0.9425 \\
26 & {\tt NLP Group 7} & 0.6499 & 0.8462 & 0.8605 & 0.9276 \\
27 & {\tt NLIP} & 0.6438 & 0.8546 & 0.8723 & 0.9257 \\
28 & {\tt ALA} & 0.6361 & 0.8423 & 0.8493 & 0.9140 \\
29 & {\tt mucai} & 0.6285 & 0.8067 & 0.8354 & 0.8985 \\
30 & {\tt AGS} & 0.6251 & 0.8390 & 0.8640 & 0.9211 \\
31 & {\tt Tutorify} & 0.6247 & 0.8261 & 0.8502 & 0.9173 \\
32 & {\tt Promptly Educated} & 0.6196 & 0.7104 & 0.8479 & 0.9224 \\
33 & {\tt wyn} & 0.6184 & 0.8384 & 0.8434 & 0.9095 \\
34 & {\tt Emergent Wisdom} & 0.6100 & 0.8546 & 0.8799 & 0.9321 \\
35 & {\tt Georgia Tech EDU} & 0.6049 & 0.8171 & 0.8386 & 0.9102 \\
36 & {\tt SG} & 0.5896 & 0.7919 & 0.8258 & 0.8875 \\
37 & {\tt NeuralNexus} & 0.5840 & 0.8268 & 0.8142 & 0.8972 \\
38 & {\tt presidency} & 0.5807 & 0.7570 & 0.8070 & 0.8804 \\
39 & {\tt NLP\_UNH} & 0.5708 & 0.8358 & 0.8358 & 0.9089 \\
40 & {\tt letstea} & 0.5376 & 0.6593 & 0.8109 & 0.8681 \\
41 & {\tt Patriots} & 0.5345 & 0.8028 & 0.7923 & 0.8921 \\
42 & {\tt AUST\_NLP} & 0.4819 & 0.7085 & 0.6929 & 0.7576 \\
43 & {\tt WhyIamHere} & 0.4562 & 0.7931 & 0.7126 & 0.8824 \\
44 & {\tt RAGthoven} &	0.2949 & 0.4350 & 0.4349 & 0.5365 \\
\hline
\end{tabular}
}
\caption{Official leaderboard for {\em Track 1: Mistake Identification}}
\label{tab:lead-MI}
\end{table*}

\begin{table*}[!htb]
\centering
\small
\resizebox{0.8\linewidth}{!}{
\begin{tabular}{|p{0.1\linewidth}|p{0.4\linewidth}|p{0.1\linewidth}|}
\hline
\textbf{Rank} & \textbf{Team} & \textbf{Ex. Acc} \\ 
\hline
1 & {\tt TutorMind} & 0.8798 \\
  & {\tt MSA} & \\
2 & {\tt BD} & 0.8772 \\ 
3 & {\tt BJTU} & 0.8765 \\
4 & {\tt Archaeology} & 0.8746 \\
5 & {\tt TBA} & 0.8740 \\
\hline
\end{tabular}
}
\caption{Top 5 results according to exact accuracy for {\em Track 1: Mistake Identification}}
\label{tab:lead-MI-2}
\end{table*}

\begin{table*}[!htb]
\small
\centering
\resizebox{0.8\linewidth}{!}{
\begin{tabular}{|p{0.1\linewidth}|p{0.4\linewidth}|p{0.1\linewidth}|}
\hline
\textbf{Rank} & \textbf{Team} & \textbf{Len. F1} \\ 
\hline
1 & {\tt BJTU} & 0.9185\\
2 & {\tt MSA} & 0.9152\\
3 & {\tt TutorMind} & 0.9143 \\
4 & {\tt BLCU-ICALL} & 0.9110 \\
5 & {\tt Wonderland\_EDU@HKU} & 0.9109 \\ 
\hline
\end{tabular}
}
\caption{Top 5 results according to lenient F1 for {\em Track 1: Mistake Identification}}
\label{tab:lead-MI-3}
\end{table*}

\begin{table*}[!htb]
\small
\centering
\resizebox{0.8\linewidth}{!}{
\begin{tabular}{|p{0.1\linewidth}|p{0.4\linewidth}|p{0.1\linewidth}|}
\hline
\textbf{Rank} & \textbf{Team} & \textbf{Len. Acc} \\ 
\hline
1 & {\tt TutorMind} & 0.9541 \\
2 & {\tt MSA} & 0.9535 \\
  & {\tt BJTU} & \\
3 & {\tt BLCU-ICALL} & 0.9515 \\
4 & {\tt Wonderland\_EDU@HKU} & 0.9496 \\
5 & {\tt TBA} & 0.9476 \\
\hline
\end{tabular}
}
\caption{Top 5 results according to lenient accuracy for {\em Track 1: Mistake Identification}}
\label{tab:lead-MI-4}
\end{table*}

%%%%%%%%%%%%%

%\subsection{Track 2: Mistake Location}

%The main leaderboard is presented by Table \ref{tab:lead-ML}. The top 5 results for each secondary metric in this track can be found in Tables \ref{tab:lead-ML-2} to \ref{tab:lead-ML-4}.

\begin{table*}[!htb]
\centering
\resizebox{\linewidth}{!}{
\begin{tabular}{|p{0.1\linewidth}|p{0.4\linewidth}|p{0.1\linewidth}|p{0.1\linewidth}|p{0.1\linewidth}|p{0.1\linewidth}|}
\hline
\textbf{Rank} & \textbf{Team} & \textbf{Ex. F1} & \textbf{Ex. Acc} & \textbf{Len. F1} & \textbf{Len. Acc} \\ 
\hline
1 & {\tt BLCU-ICALL} & 0.5983 & 0.7679 & 0.8386 & 0.8630 \\
2 & {\tt BJTU} & 0.5940 & 0.7330 & 0.7848 & 0.8261 \\
3 & {\tt K-NLPers} & 0.5880 & 0.7641 & 0.8404 & 0.8610 \\
4 & {\tt MSA} & 0.5743 & 0.6975 & 0.7848 & 0.8209 \\
5 & {\tt SG} & 0.5692 & 0.7602 & 0.8118 & 0.8416 \\
6 & {\tt bea-jh} & 0.5658 & 0.6723 & 0.7792 & 0.8197 \\
7 & {\tt bd} & 0.5543 & 0.7143 & 0.7699 & 0.8054 \\
8 & {\tt TBA} & 0.5490 & 0.7091 & 0.7702 & 0.8035 \\
9 & {\tt Wonderland\_EDU@HKU} & 0.5450 & 0.7104 & 0.7649 & 0.8003 \\
10 & {\tt Averroes} & 0.5366 & 0.6348 & 0.7587 & 0.7822 \\
11 & {\tt Whyamher} & 0.5325 & 0.6910 & 0.7370 & 0.7802 \\
12 & {\tt NLIP} & 0.5319 & 0.6878 & 0.7495 & 0.7951 \\
   & {\tt Archaeology} & 0.5319 & 0.6568 & 0.7558 & 0.8009 \\
13 & {\tt JiNan\_Smart Education} & 0.5274 & 0.6968 & 0.7502 & 0.7809 \\
14 & {\tt Squirrel Ai Learning} & 	0.5272 & 0.6904 & 0.7306 & 0.7692 \\
15 & {\tt Thapar Titans} & 0.5215 & 0.6943 & 0.7374 & 0.7796 \\
16 & {\tt jeez} & 0.5187 & 0.6833 & 0.7416 & 0.7854 \\
17 & {\tt CU} & 0.5148 & 0.6807 & 0.7358 & 0.7789 \\
18 & {\tt 777} & 0.5114 & 0.6710 & 0.7195 & 0.7486 \\
19 & {\tt Someone} & 0.5009 & 0.7208 & 0.7590 & 0.8074 \\
20 & {\tt Retuyt-InCo} & 0.4959 & 0.5863 & 0.7200 & 0.7608 \\
21 & {\tt NLP Group 7} & 0.4936 & 0.6348 & 0.6944 & 0.7524 \\
22 & {\tt SmolLab\_SEU} & 0.4935 & 0.6057 & 0.7051 & 0.7401 \\
23 & {\tt lexilogic} & 0.4844 & 0.6548 & 0.7138 & 0.7447 \\
24 & {\tt mucai} & 0.4828 & 0.5495 & 0.7086 & 0.7343 \\
25 & {\tt Emergent Wisdom} & 0.4773 & 0.7188 & 0.7436 & 0.7893 \\
26 & {\tt 2} & 0.4749 & 0.7279 & 0.7397 & 0.8003 \\
27 & {\tt Promptly Educated} & 0.4717 & 0.6432 & 0.6900 & 0.7337 \\
28 & {\tt Tutorify} & 0.4666 & 0.6626 & 0.7116 & 0.7447 \\
29 & {\tt NLP\_UNH} & 0.4515 & 0.6994 & 0.6962 & 0.7725 \\
30 & {\tt Patriots} & 0.4450 & 0.6328 & 0.6548 & 0.7007 \\
31 & {\tt AUST\_NLP} & 0.3044 & 0.4163 & 0.4759 & 0.4848 \\
\hline
\end{tabular}
}
\caption{Official leaderboard for {\em Track 2: Mistake Location}}
\label{tab:lead-ML}
\end{table*}

\begin{table*}[!htb]
\centering
\small
\resizebox{0.8\linewidth}{!}{
\begin{tabular}{|p{0.1\linewidth}|p{0.4\linewidth}|p{0.1\linewidth}|}
\hline
\textbf{Rank} & \textbf{Team} & \textbf{Ex. Acc} \\ 
\hline
1 & {\tt BLCU-ICALL} & 0.7679 \\
2 & {\tt K-NLPers} & 0.7641 \\
3 & {\tt SG} & 0.7602 \\
4 & {\tt bea-jh} & 0.7389 \\
5 & {\tt BJTU} & 0.7330 \\
\hline
\end{tabular}
}
\caption{Top 5 results according to exact accuracy for {\em Track 2: Mistake Location}}
\label{tab:lead-ML-2}
\end{table*}

\begin{table*}[!htb]
\small
\centering
\resizebox{0.8\linewidth}{!}{
\begin{tabular}{|p{0.1\linewidth}|p{0.4\linewidth}|p{0.1\linewidth}|}
\hline
\textbf{Rank} & \textbf{Team} & \textbf{Len. F1} \\ 
\hline
1 & {\tt K-NLPers} & 0.8404 \\
2 & {\tt BLCU-ICALL} & 0.8386 \\
3 & {\tt SG} & 0.8118 \\
4 & {\tt BJTU} & 0.7861 \\
5 & {\tt bea-jh} & 0.7851 \\ 
\hline
\end{tabular}
}
\caption{Top 5 results according to lenient F1 for {\em Track 2: Mistake Location}}
\label{tab:lead-ML-3}
\end{table*}

\begin{table*}[!htb]
\small
\centering
\resizebox{0.8\linewidth}{!}{
\begin{tabular}{|p{0.1\linewidth}|p{0.4\linewidth}|p{0.1\linewidth}|}
\hline
\textbf{Rank} & \textbf{Team} & \textbf{Len. Acc} \\ 
\hline
1 & {\tt BLCU-ICALL} & 0.8630 \\
2 & {\tt K-NLPers} & 0.8610 \\
3 & {\tt SG} & 0.8416 \\
4 & {\tt BJTU} & 0.8274 \\
5 & {\tt bea-jh} & 0.8268 \\
\hline
\end{tabular}
}
\caption{Top 5 results according to lenient accuracy for {\em Track 2: Mistake Location}}
\label{tab:lead-ML-4}
\end{table*}

%%%%%%%%%%%
%\subsection{Track 3: Providing Guidance}

%Table \ref{tab:lead-PG} presents the main leaderboard, while the top 5 results for each secondary metric in this track can be found in Tables \ref{tab:lead-PG-2} to \ref{tab:lead-PG-4}.

\begin{table*}[!htb]
\centering
\resizebox{\linewidth}{!}{
\begin{tabular}{|p{0.1\linewidth}|p{0.4\linewidth}|p{0.1\linewidth}|p{0.1\linewidth}|p{0.1\linewidth}|p{0.1\linewidth}|}
\hline
\textbf{Rank} & \textbf{Team} & \textbf{Ex. F1} & \textbf{Ex. Acc} & \textbf{Len. F1} & \textbf{Len. Acc} \\ 
\hline
1 & {\tt MSA} & 0.5834 & 0.6613 & 0.7798 & 0.8190 \\
2 & {\tt SG} & 0.5785 & 0.7052 & 0.7860 & 0.8216 \\
3 & {\tt BLCU-ICALL} & 0.5741 & 0.6716 & 0.7487 & 0.8061 \\
4 & {\tt BJTU} & 0.5725 & 0.6490 & 0.7445 & 0.8100 \\
5 & {\tt K-NLPers} & 0.5606 & 0.6270 & 0.7446 & 0.8003 \\
6 & {\tt bea-jh} & 0.5451 & 0.6387 & 0.7253 & 0.7977 \\
7 & {\tt Wonderland\_EDU@HKU} & 0.5416 & 0.6464 & 0.7456 & 0.7886 \\
8 & {\tt IALab UC} & 0.5369 & 0.6244 & 0.7379 & 0.7822 \\
9 & {\tt JiNan\_Smart Education} & 0.5275 & 0.6432 & 0.7336 & 0.7893 \\
10 & {\tt Henry} & 0.5265 & 0.6238 & 0.7196 & 0.7744 \\
11 & {\tt TBA} & 0.5212 & 0.6219 & 0.7299 & 0.7906 \\
12 & {\tt MT-NLP} & 0.5211 & 0.6141 & 0.7142 & 0.7699 \\
13 & {\tt Archaeology} & 0.5208 & 0.5734 & 0.7171 & 0.7770 \\
14 & {\tt Averroes} & 0.5134 & 0.6309 & 0.7095 & 0.7751 \\
15 & {\tt Squirrel Ai Learning} & 0.5087 & 0.6005 & 0.7059 & 0.7763 \\
16 & {\tt jeez} & 0.5071 & 0.5831 & 0.7234 & 0.7763 \\
   & {\tt bd} & 0.5071 & 0.5831 & 0.7234 & 0.7763 \\
17 & {\tt Retuyt-InCo} & 0.5049 & 0.5947 & 0.7057 & 0.7751 \\
18 & {\tt woaiyuanshen} & 0.4974 & 0.5798 & 0.7034 & 0.7841 \\
19 & {\tt SmolLab\_SEU} & 0.4933 & 0.5695 & 0.6990 & 0.7608 \\
20 & {\tt CU} & 0.4926 & 0.5850 & 0.7031 & 0.7692 \\
21 & {\tt Emergent Wisdom} & 0.4903 & 0.6102 & 0.6919 & 0.7725 \\
22 & {\tt NLIP} & 0.4888 & 0.6025 & 0.6927 & 0.7647 \\
23 & {\tt batikbabu} & 0.4873 & 0.6147 & 0.7001 & 0.7615 \\
24 & {\tt Whyiamhere} & 0.4856 & 0.6231 & 0.6880 & 0.7738 \\
25 & {\tt isistanNiem} & 0.4805 & 0.5844 & 0.6715 & 0.7589 \\
26 & {\tt Thapar Titans} & 0.4777 & 0.5624 & 0.6846 & 0.7479 \\
27 & {\tt DLSU} & 0.4776 & 0.5669 & 0.6755 & 0.7382 \\
28 & {\tt Tutorify} & 0.4731 & 0.5753 & 0.6709 & 0.7511 \\
29 & {\tt 777} & 0.4711 & 0.6432 & 0.7075 & 0.7725 \\
30 & {\tt Promptly Educated} & 0.4674 & 0.6102 & 0.6785 & 0.7647 \\
31 & {\tt lexiLogic} & 0.4656 & 0.5869 & 0.6803 & 0.7473 \\
32 & {\tt GGEZ Lab} & 0.4596 & 0.5714 & 0.6652 & 0.7492 \\
33 & {\tt Patriots} & 0.4508 & 0.5663 & 0.6422 & 0.7311 \\
34 & {\tt NLP\_UNH} & 0.4301 & 0.6380 & 0.6895 & 0.7692 \\
35 & {\tt AUST\_NLP} & 0.4045 & 0.5973 & 0.6094 & 0.7259 \\
\hline
\end{tabular}
}
\caption{Official leaderboard for {\em Track 3: Providing Guidance}}
\label{tab:lead-PG}
\end{table*}

\begin{table*}[!htb]
\centering
\small
\resizebox{0.8\linewidth}{!}{
\begin{tabular}{|p{0.1\linewidth}|p{0.4\linewidth}|p{0.1\linewidth}|}
\hline
\textbf{Rank} & \textbf{Team} & \textbf{Ex. Acc} \\ 
\hline
1 & {\tt SG} & 0.7052 \\
2 & {\tt BLCU-ICALL} & 0.7007 \\
3 & {\tt MSA} & 0.6729 \\
4 & {\tt bea-jh} & 0.6703 \\
5 & {\tt TBA} & 0.6652 \\
\hline
\end{tabular}
}
\caption{Top 5 results according to exact accuracy for {\em Track 3: Providing Guidance}}
\label{tab:lead-PG-2}
\end{table*}

\begin{table*}[!htb]
\small
\centering
\resizebox{0.8\linewidth}{!}{
\begin{tabular}{|p{0.1\linewidth}|p{0.4\linewidth}|p{0.1\linewidth}|}
\hline
\textbf{Rank} & \textbf{Team} & \textbf{Len. F1} \\ 
\hline
1 & {\tt SG} & 0.7860 \\
2 & {\tt MSA} & 0.7798 \\
3 & {\tt BLCU-ICALL} & 0.7699 \\
4 & {\tt K-NLPers} & 0.7483 \\
5 & {\tt Wonderland\_EDU@HKU} & 0.7456\\ 
\hline
\end{tabular}
}
\caption{Top 5 results according to lenient F1 for {\em Track 3: Providing Guidance}}
\label{tab:lead-PG-3}
\end{table*}

\begin{table*}[!htb]
\small
\centering
\resizebox{0.8\linewidth}{!}{
\begin{tabular}{|p{0.1\linewidth}|p{0.4\linewidth}|p{0.1\linewidth}|}
\hline
\textbf{Rank} & \textbf{Team} & \textbf{Len. Acc} \\ 
\hline
1 & {\tt BLCU-ICALL} & 0.8222 \\
2 & {\tt SG} & 0.8216 \\
3 & {\tt MSA} & 0.8190 \\
4 & {\tt BJTU} & 0.8100 \\
5 & {\tt TBA} & 0.8035 \\
\hline
\end{tabular}
}
\caption{Top 5 results according to lenient accuracy for {\em Track 3: Providing Guidance}}
\label{tab:lead-PG-4}
\end{table*}

%%%%%%%%%%%
%\subsection{Track 4: Actionability}

%Table \ref{tab:lead-AC} presents the main leaderboard, while Tables \ref{tab:lead-AC-2} to \ref{tab:lead-AC-4} report on the top 5 results for each secondary metric in this track.

\begin{table*}[!htb]
\centering
\resizebox{\linewidth}{!}{
\begin{tabular}{|p{0.1\linewidth}|p{0.4\linewidth}|p{0.1\linewidth}|p{0.1\linewidth}|p{0.1\linewidth}|p{0.1\linewidth}|}
\hline
\textbf{Rank} & \textbf{Team} & \textbf{Ex. F1} & \textbf{Ex. Acc} & \textbf{Len. F1} & \textbf{Len. Acc} \\ 
\hline
1 & {\tt bea-jh} & 0.7085 & 0.7298 & 0.8527 & 0.8837 \\
2 & {\tt BJTU} & 0.6992 & 0.7363 & 0.8633 & 0.8940 \\
3 & {\tt MSA} & 0.6984 & 0.7537 & 0.8659 & 0.8908 \\
4 & {\tt lexiLogic} & 0.6930 & 0.7162 & 0.8393 & 0.8675 \\
5 & {\tt Phaedrus} & 0.6907 & 0.7298 & 0.8346 & 0.8656 \\
6 & {\tt Wonderland\_EDU@HKU} & 0.6843 & 0.7285 & 0.8613 & 0.8888 \\
7 & {\tt Archaeology} & 0.6776 & 0.7214 & 0.8302 & 0.8565 \\
8 & {\tt BLCU-ICALL} & 0.6735 & 0.7363 & 0.8596 & 0.8856 \\
9 & {\tt TBA} & 0.6671 & 0.7324 & 0.8499 & 0.8752 \\
10 & {\tt 4} & 0.6668 & 0.7033 & 0.8160 & 0.8520 \\
   & {\tt JiNan\_Smart Education} & 0.6668 & 0.7033 & 0.8160 & 0.8520 \\
11 & {\tt bnl} & 0.6655 & 0.6813 & 0.8172 & 0.8597 \\
12 & {\tt woaiyuanshen} & 0.6651 & 0.7124 & 0.8191 & 0.8533 \\
13 & {\tt love-peace} & 0.6562 & 0.6839 & 0.8051 & 0.8352 \\
14 & {\tt bd} & 0.6554 & 0.7182 & 0.8461 & 0.8707 \\
15 & {\tt Thapar Titans} & 0.6324 & 0.6774 & 0.7936 & 0.8248 \\
16 & {\tt SmolLab\_SEU} & 0.6284 & 0.6955 & 0.8223 & 0.8565 \\
17 & {\tt Retuyt-InCo} & 0.6129 & 0.7033 & 0.8272 & 0.8559 \\
18 & {\tt NLIP} & 0.6055 & 0.6897 & 0.8205 & 0.8468 \\
19 & {\tt Squirrel Ai Learning} & 0.5954 & 0.6516 & 0.7639 & 0.8022 \\
20 & {\tt Tutorify} & 0.5681 & 0.6425 & 0.7749 & 0.8190 \\
21 & {\tt K-NLPers} & 0.5664 & 0.5773 & 0.7346 & 0.8061 \\
22 & {\tt Emergent Wisdom} & 0.5661 & 0.6645 & 0.7782 & 0.8054 \\
23 & {\tt SG} & 0.5465 & 0.6341 & 0.7545 & 0.7725 \\
24 & {\tt SAI} & 0.5398 & 0.6277 & 0.7564 & 0.8022 \\
25 & {\tt DLSU} & 0.5294 & 0.6089 & 0.7351 & 0.7738 \\
26 & {\tt Patriots} & 0.4630 & 0.5727 & 0.6943 & 0.7537 \\
27 & {\tt whyiamhere} & 0.4306 & 0.6044 & 0.7143 & 0.7938 \\
28 & {\tt AUST\_NLP} & 0.4196 & 0.5262 & 0.6077 & 0.6833 \\
29 & {\tt NLP\_UNH} & 0.3798 & 0.5546 & 0.6530 & 0.7524 \\
\hline
\end{tabular}
}
\caption{Official leaderboard for {\em Track 4: Actionability}}
\label{tab:lead-AC}
\end{table*}

\begin{table*}[!htb]
\centering
\small
\resizebox{0.8\linewidth}{!}{
\begin{tabular}{|p{0.1\linewidth}|p{0.4\linewidth}|p{0.1\linewidth}|}
\hline
\textbf{Rank} & \textbf{Team} & \textbf{Ex. Acc} \\ 
\hline
1 & {\tt bea-jh} & 0.7557 \\
2 & {\tt MSA} & 0.7537 \\
3 & {\tt BJTU} & 0.7363 \\
  & {\tt BLCU-ICALL} & \\
4 & {\tt TBA} & 0.7324 \\
5 & {\tt Phaedrus} & 0.7298 \\
\hline
\end{tabular}
}
\caption{Top 5 results according to exact accuracy for {\em Track 4: Actionability}}
\label{tab:lead-AC-2}
\end{table*}

\begin{table*}[!htb]
\small
\centering
\resizebox{0.8\linewidth}{!}{
\begin{tabular}{|p{0.1\linewidth}|p{0.4\linewidth}|p{0.1\linewidth}|}
\hline
\textbf{Rank} & \textbf{Team} & \textbf{Len. F1} \\ 
\hline
1 & {\tt MSA} & 0.8659 \\
2 & {\tt BJTU} & 0.8633 \\
3 & {\tt Wonderland\_EDU@HKU} & 0.8613 \\
4 & {\tt bea-jh} & 0.8609 \\
5 & {\tt BLCU-ICALL} & 0.8596\\ 
\hline
\end{tabular}
}
\caption{Top 5 results according to lenient F1 for {\em Track 4: Actionability}}
\label{tab:lead-AC-3}
\end{table*}

\begin{table*}[!htb]
\small
\centering
\resizebox{0.8\linewidth}{!}{
\begin{tabular}{|p{0.1\linewidth}|p{0.4\linewidth}|p{0.1\linewidth}|}
\hline
\textbf{Rank} & \textbf{Team} & \textbf{Len. Acc} \\ 
\hline
1 & {\tt BJTU} & 0.8940 \\
2 & {\tt MSA} & 0.8908 \\
3 & {\tt Wonderland\_EDU@HKU} & 0.8888 \\
4 & {\tt bea-jh} & 0.8875 \\
5 & {\tt BLCU-ICALL} & 0.8856\\
\hline
\end{tabular}
}
\caption{Top 5 results according to lenient accuracy for {\em Track 4: Actionability}}
\label{tab:lead-AC-4}
\end{table*}

%%%%%%%%%%%
%\subsection{Track 5: Tutor Identification}

%Tables \ref{tab:lead-TI} and \ref{tab:lead-TI-2} present the main leaderboard and the top 5 results for the secondary metric, respectively.

\begin{table*}[!htb]
\centering
\resizebox{0.8\linewidth}{!}{
\begin{tabular}{|p{0.1\linewidth}|p{0.4\linewidth}|p{0.1\linewidth}|p{0.1\linewidth}|}
\hline
\textbf{Rank} & \textbf{Team} & \textbf{Ex. F1} & \textbf{Ex. Acc} \\ 
\hline
1 & {\tt Phaedrus} & 0.9698 & 0.9664 \\
2 & {\tt SYSUpporter} & 0.9692 & 0.9657 \\
3 & {\tt Two Outliers} & 0.9172 & 0.9412 \\
4 & {\tt JInan\_Smart Education} & 0.8965 & 0.8940 \\
5 & {\tt BLCU-ICALL} & 0.8930 & 0.8908 \\
6 & {\tt Archaeology} & 0.8866 & 0.8882 \\
7 & {\tt Wonderland\_EDU@HKU} & 0.8795 & 0.8778 \\
8 & {\tt MSA} & 0.8697 & 0.8649 \\
9 & {\tt SmolLab\_SEU} & 0.8621 & 0.8604 \\
10 & {\tt mucai} & 0.8602 & 0.8675 \\
11 & {\tt Squirrel Ai Learning} & 0.8432 & 0.8390 \\
12 & {\tt Retuyt-InCo} & 0.8385 & 0.8475 \\
13 & {\tt whyiamhere} & 0.8356 & 0.8345 \\
14 & {\tt bnl} & 0.8247 & 0.8216 \\
15 & {\tt Tutorify} & 0.8212 & 0.8100 \\
16 & {\tt LexiLogic} & 0.8207 & 0.8145 \\
17 & {\tt Georgia Tech EDU} & 0.6468 & 0.6296 \\
18 & {\tt DLSU} & 0.6420 & 0.6231 \\
19 & {\tt letstea} & 0.1749 & 0.1635 \\
20 & {\tt zet-epsilon} & 0.1140 & 0.1965 \\
\hline
\end{tabular}
}
\caption{Official leaderboard for {\em Track 5: Tutor Identification}}
\label{tab:lead-TI}
\end{table*}

\begin{table*}[!t]
\centering
\small
\resizebox{0.8\linewidth}{!}{
\begin{tabular}{|p{0.1\linewidth}|p{0.4\linewidth}|p{0.1\linewidth}|}
\hline
\textbf{Rank} & \textbf{Team} & \textbf{Ex. Acc} \\ 
\hline
1 & {\tt Phaedrus} & 0.9664 \\
2 & {\tt SYSUpporter} & 0.9657 \\
3 & {\tt Two Outliers} & 0.9412 \\
4 & {\tt JInan\_Smart Education} & 0.8940 \\
5 & {\tt BLCU-ICALL} & 0.8908 \\
\hline
\end{tabular}
}
\caption{Top 5 results according to exact accuracy for {\em Track 5: Tutor Identification}}
\label{tab:lead-TI-2}
\end{table*}

\onecolumn

\newpage

%%%%%%%%
%%%%%%%%
%\section{Analysis of the approaches}
%\label{app:approaches}
%Include a table overviewing the major trends per track -- what types of models did teams use?

%Table \ref{tab:teamsummary} provides a comprehensive overview of the modeling approaches and LLMs employed by participating teams. Table \ref{tab:top5_teams} and Figure \ref{fig:topteams} further summarize the methodologies and models adopted by the top-performing submissions in each track. Notably, Table \ref{tab:top5_teams} highlights the highest-scoring teams across tracks. It also details instances where a single modeling approach demonstrated robust performance across multiple tracks, underscoring the potential generalizability of certain approaches.

% \begin{itemize}
%    \item team leaderboards (with submissions pruned, just one best result per team within the first 5 submitted by the team in case they submitted more than 5). The ranking metric is the main one for the shared task -- exact macro F1.
%    \item smaller leaderboards -- one per metric (exact F1, exact acc, lenient F1, lenient acc), with the appropriate ranking per metric. Here, the top 5 best results would be enough. These can come from any of the 5 allowed submissions per team (e.g., Team 1 may win according to the macro F1 from their submitted System 1, and they may win according to accuracy from their submitted System 2).
%     \item Include a table overviewing the major trends per track -- what types of models did teams use?
% \end{itemize}

% TODO (Kaushal): Add summary table for all teams
\begin{table*}[!h]
\centering
\resizebox{\textwidth}{!}{
\begin{tabular}{p{5cm}|p{8cm}|p{7cm}}
\hline \hline
\textbf{Team} & \textbf{Keywords of the Approach} & \textbf{Models / LLMs} \\
\hline \hline
BJTU \cite{fan-2025-shared} & Zero-shot prompting, data augmentation, task-aware prompting & Unspecified \\ \hline
TutorMind \cite{dekmak-2025-shared} & Instruction tuning (LoRA), data augmentation & \texttt{GPT-4o-mini}, {\tt LLaMA-3.1-8B}, \texttt{Mistral-7B} \\ \hline
Averroes \cite{yasser-2025-shared} & Instruction tuning (LoRA) & {\tt GTE-ModernBERT-Base}, \texttt{GTE-Qwen2-1.5B}, \texttt{Qwen2.5-0.5B}, \texttt{Qwen2.5-1.5B}, \texttt{Qwen2.5-Math-1.5B} \\ \hline
MSA \cite{hikal-2025-shared} & Instruction tuning (LoRA), ensembling & {\tt Mathstral-7B-v0.1} \\ \hline
BD \cite{rohan-2025-shared} & SFT, class weighting, ensembling & \texttt{MPNet} \\ \hline
Wonderland\_EDU@HKU \cite{deliang-2025-shared} & Instruction tuning (LoRA) & \texttt{LLaMA-3.2-3B} \\ \hline
Archaeology \cite{roșu-2025-shared} & SFT, fine-tuning (LoRA), binary-to-multiclass phase-wise training, fitting statistical classifiers & \texttt{Logistic Regression}, \texttt{LightGBM}, \texttt{String-Kernel-SVM}, \texttt{RoBERTa}, \texttt{DeBERTa}, \texttt{ModernBERT}, \texttt{GritLM}, \texttt{GPT2-XL}, \texttt{Mistral-7B}, \texttt{XGBoost} \\ \hline
TBA \cite{gombert-2025-shared} & SFT, DARE-TIES algorithm & \texttt{FLAN-T5-XL} \\ \hline
BLCU-ICALL \cite{an-2025-shared} & SFT, few-shot prompting, RLHF & {\tt GPT-4o}, \texttt{GPT-o3-mini}, \texttt{Gemini-2.5-pro}, \texttt{Grok-3}, \texttt{Deepseek-R1}, \texttt{Claude-3.7}, \texttt{LLaMA-3.1-8B}, \texttt{QwQ-32B}, \texttt{Qwen2.5-7B}, \texttt{Qwen2.5-14B}, \texttt{Qwen2.5-32B} \\ \hline
bea-jh \cite{roh-2025-shared} & GRPO, thinking-based model & \texttt{GLM-4-9B} \\ \hline
JiNan\_Smart Education \cite{chen-2025-shared} & Input representation, representation fusion, ensembling & \texttt{DeBERTa-v3} \\ \hline
K-NLPers \cite{park-2025-shared} & Chain-of-thought prompting, multi-strategy ensembling, input representation & \texttt{GPT-4.1} \\ \hline
Thapar Titan/s \cite{dadwal-2025-shared} & Data augmentation, weighted loss, SFT & \texttt{BERT}, \texttt{DeBERTa}, \texttt{RoBERTa} \\ \hline
SmolLab\_SEU \cite{rahman-2025-shared} & SFT & \texttt{DeBERTa-V3}, \texttt{EduBERT}, \texttt{RoBERTa-Large}, \texttt{SciBERT} \\ \hline
LexiLogic \cite{bhattacharyya-2025-shared} & SFT, zero-shot prompting, few-shot prompting & \texttt{Flan-T5}, \texttt{Llama-3.2-3B}, \texttt{Llama-3-8B}, \texttt{ModernBERT}, \texttt{MathBERT}, \texttt{Phi-4-mini-instruct}, \texttt{Qwen2.5-7B-Instruct} \\ \hline
Retuyt-InCo \cite{gongora-2025-shared} & Input representation, SFT, fitting statistical classifiers & \texttt{Random Forest}, \texttt{SVC}, \texttt{k-NN}, \texttt{Qwen2.5-0.5B-Instruct}, \texttt{XGBoost}, \texttt{DistilBERT}, \texttt{BERT} \\ \hline
CU \cite{zhihao-2025-shared} & SFT, data augmentation, label imbalance mitigation & \texttt{BERT}, \texttt{GPT-4.1} \\ \hline
NLIP \cite{saha-2025-shared} & SFT, data augmentation, multi-task learning, ensembling & \texttt{RoBERTa}, \texttt{DeBERTa} \\ \hline
ALA* & SFT & \texttt{BERT} \\ \hline
Emergent Wisdom \cite{jain-2025-shared} & Input representation, feature engineering, routing, fitting statistical classifiers, multi-step prompting, LLM judge & \texttt{XGBoost}, \texttt{T5} \\ \hline
SG* & Multi-step prompting & \texttt{Gemma-3-27B-IT} \\ \hline
NeuralNexus \cite{naeem-2025-shared} & RAG, few-shot prompting, fitting statistical classifiers & \texttt{k-NN}, \texttt{GPT-4o} \\ \hline
IALab UC \cite{busquets-2025-shared} & Zero-shot prompting, feature engineering, pedagogical theory, fitting statistical classifiers & \texttt{LearnLM-1.5}, \texttt{Random Forest} \\ \hline
Henry \cite{pit-2025-shared} & Zero-shot prompting, GRPO,  fine-tuning (LoRA), modified reward optimization & \texttt{GPT-4o}, \texttt{Claude 2.7 Sonnet}, \texttt{Phi-3.5-mini Instruct}, \texttt{MLP} \\ \hline
DLSU \cite{manlises-2025-shared} & Input representation, fitting statistical classifiers & \texttt{gte-modernbert-base}, \texttt{all-MiniLM-L12-v2}, \texttt{MLP} \\ \hline
Phaedrus \cite{tiwari-2025-shared} & Zero-shot prompting, instruction tuning (LoRA), ensembling & \texttt{DeBERTa-v3-large}, \texttt{DeBERTa-v3-base}, \texttt{DeBERTa-v3-small}, \texttt{Longformer-base-4096}, \texttt{BigBird-RoBERTa-large}, \texttt{Qwen-2.5-0.5B}, \texttt{Zephyr-7B-alpha} \\ \hline
SYSUpporter \cite{chen-huang-2025-shared} & Data augmentation, class weighting, ensembling, Hungarian algorithm & \texttt{Logistic Regression}, \texttt{Random Forest}, \texttt{Extra Trees}, \texttt{XGBoost}, \texttt{DeBERTa} \\ \hline
Two Outliers \cite{tjitrahardja-2025-shared} & Input representation, supervised contrastive learning, ensembling, CalBoost, Hungarian algorithm & \texttt{DeBERTa} \\ \hline
Gooby-Snoob Guysz* & Prompt optimization,  failure-driven prompting & \texttt{OpenAI's O1}, \texttt{GPT-4o} \\ \hline\hline
\end{tabular}
}
\caption{ \small 
Keywords and models associated with the approaches adopted by participating teams across all tracks. SFT = Supervised Fine-Tuning, 
RAG = Retrieval-Augmented Generation, 
RLHF = Reinforcement Learning from Human Feedback, 
MLP = Multilayer Perceptron, 
GRPO = Guided Reward Prompt Optimization. *Details are obtained via email correspondence. 
\textit{Statistical classifiers} include traditional models such as Random Forest (RF), XGBoost, etc.
}
\label{tab:teamsummary}
\end{table*}

% TODO (Kaushal): Add summary table for top perfoming teams
\begin{table*}[!h]
    \centering
    \resizebox{\textwidth}{!}{
    \begin{tabular}{l|p{5cm}|p{8cm}|p{7cm}}
        \hline \hline
        \textbf{Track/Criteria} & \textbf{Teams} & \textbf{Keywords for Approaches} & \textbf{Model/LLMs} \\
        \hline \hline
        *Mistake Identification & \texttt{BJTU}, \texttt{Tutor Mind}, \texttt{Averroes}, \texttt{BD}, \texttt{MSA}  &  Zero-shot prompting, data augmentation, task-aware prompting, instruction tuning (LoRA), ensembling, SFT, class weighting & \texttt{GPT-4o-mini}, \texttt{LLaMA-3.1-8B}, \texttt{Mistral-7B}, \texttt{GTE-ModernBERT-Base}, \texttt{GTE-Qwen2-1.5B}, 
\texttt{Qwen2.5-0.5B}, \texttt{Qwen2.5-1.5B}, 
\texttt{Qwen2.5-Math-1.5B}, \texttt{Mathstral-7B-v0.1}, \texttt{MPNet}
 \\ 
        \hline
        *Mistake Location & \texttt{BLCU-ICALL}, \texttt{BJTU}, \texttt{K-NLPers}, \texttt{MSA}, \texttt{SG} & SFT, few-shot prompting, RLHF, zero-shot prompting, data augmentation, task-aware prompting, chain-of-thought prompting, multi-step prompting, multi-strategy ensembling, input representation & \texttt{GPT-4o}, \texttt{GPT-o3-mini}, \texttt{Gemini-2.5-pro}, 
\texttt{Grok-3}, \texttt{Deepseek-R1}, \texttt{Claude-3.7}, 
\texttt{LLaMA-3.1-8B}, \texttt{QwQ-32B},
\texttt{Qwen2.5-7B}, \texttt{Qwen2.5-14B}, \texttt{Qwen2.5-32B}, 
\texttt{GPT-4.1}, \texttt{Mathstral-7B-v0.1}, \texttt{Gemma-3-27B-IT}
\\ 
        \hline
        *Providing Guidance & \texttt{MSA}, \texttt{SG}, \texttt{BLCU-ICALL}, \texttt{BJTU}, \texttt{K-NLPers}  & SFT, few-shot prompting, RLHF, zero-shot prompting, data augmentation, task-aware prompting, chain-of-thought prompting, multi-step prompting, multi-strategy ensembling, input representation & \texttt{GPT-4o}, \texttt{GPT-o3-mini}, \texttt{Gemini-2.5-pro}, 
\texttt{Grok-3}, \texttt{Deepseek-R1}, \texttt{Claude-3.7}, 
\texttt{LLaMA-3.1-8B}, \texttt{QwQ-32B}, 
\texttt{Qwen2.5-7B}, \texttt{Qwen2.5-14B}, \texttt{Qwen2.5-32B}, 
\texttt{GPT-4.1}, \texttt{Mathstral-7B-v0.1}, \texttt{Gemma-3-27B-IT}
 \\ 
        \hline
        *Actionability & \texttt{bea-jh}, \texttt{BJTU}, \texttt{MSA}, \texttt{lexiLogic}, \texttt{Phaedrus}  & GRPO, thinking-based model, zero-shot prompting, data augmentation, task-aware
prompting, instruction tuning (LoRA), ensembling, SFT, few-shot prompting & \texttt{GLM-4-9B}, \texttt{Mathstral-7B-v0.1}, \texttt{Flan-T5}, \texttt{ModernBERT}, \texttt{MathBERT}, \texttt{Phi-4-mini-instruct}, \texttt{Qwen2.5-7B-Instruct} \texttt{DeBERTa-v3-large}, \texttt{DeBERTa-v3-base}, \texttt{DeBERTa-v3-small}, \texttt{Longformer-base-4096}, \texttt{BigBird-RoBERTa-large}, \texttt{Qwen-2.5-0.5B}, \texttt{Zephyr-7B-alpha}
 \\ 
        \hline
        *Tutor Identification & \texttt{Phaedrus}, \texttt{SYSUpporter}, \texttt{Two Outliers}, \texttt{JInan\_Smart Education}, \texttt{BLCU-ICALL}  & SFT, few-shot prompting, RLHF, zero-shot prompting, instruction tuning (LoRA), ensembling, data augmentation, class weighting, input representation, supervised contrastive learning, CalBoost, Hungarian algorithm, representation fusion & \texttt{DeBERTa-v3-large}, \texttt{DeBERTa-v3-base}, \texttt{DeBERTa-v3-small}, \texttt{Longformer-base-4096}, \texttt{BigBird-RoBERTa-large}, \texttt{Qwen-2.5-0.5B}, \texttt{Zephyr-7B-alpha}, \texttt{GPT-4o}, \texttt{GPT-o3-mini}, \texttt{Gemini-2.5-pro}, \texttt{Grok-3}, \texttt{Deepseek-R1}, \texttt{Claude-3.7}, \texttt{LLaMA-3.1-8B}, \texttt{QwQ-32B}, \texttt{Qwen2.5-7B}, \texttt{Qwen2.5-14B}, \texttt{Qwen2.5-32B} \\ 
        \hline
        **Best Teams Across Tracks & \texttt{BJTU}, \texttt{MSA}, \texttt{BLCU-ICALL} & Zero-shot prompting, data augmentation, task-aware prompting, SFT, few-shot prompting, RLHF, instruction tuning (LoRA), ensembling  & \texttt{Mathstral-7B-v0.1}, \texttt{GPT-4o}, \texttt{GPT-o3-mini}, \texttt{Gemini-2.5-pro},  \texttt{Grok-3}, \texttt{Deepseek-R1}, \texttt{Claude-3.7},  \texttt{LLaMA-3.1-8B}, \texttt{QwQ-32B}, \texttt{Qwen2.5},  \texttt{Qwen2.5-7B}, \texttt{Qwen2.5-14B}, \texttt{Qwen2.5-32B} \\ \hline \hline

        ***Best Generalizable Models Across Tracks & \texttt{MSA}, \texttt{Wonderland\_EDU@HKU}, \texttt{TBA} & Instruction tuning (LoRA), ensembling, SFT, DARE-TIES algorithm & \texttt{Mathstral-7B-v0.1}, \texttt{LLaMA-3.2-3B}, \texttt{FLAN-T5-XL}\\ \hline \hline
    \end{tabular}
    }
    \caption{\small *Indicates Summary of top-5 team approaches and the models used per evaluation track. **Indicates teams with an average rank within the top 5 and participation in at least four tracks. ***Indicates teams that applied a generalized approach across at least four tracks and achieved an average rank of 10 or higher.}

    \label{tab:top5_teams}
\end{table*}

% \centering
%     % \resizebox{\linewidth}{!}{
%     \begin{longtable}
%     {|p{0.25\textwidth}|p{0.25\textwidth}|p{0.25\textwidth}|p{0.25\textwidth}|} \hline
%     \textbf{Team} & \textbf{Tracks} & \textbf{Approach} & \textbf{Models}\\ \hline
%     \textbf{BJTU} & 1-4 & Prompting & Unspecified\\
%     \textbf{TutorMind} & 1 & Fine-tuning & \texttt{LLaMA}, \texttt{Mistral}, \texttt{GPT-4o}\\
%     \textbf{Averroes} & 1 & Fine-tuning & \texttt{BERT}, \texttt{Qwen}\\
%     \hline
%     \end{longtable}

%%%%%%%%
%\section{Examples of Particularly Challenging Cases}
%\label{ap:difficult_examples}

\begin{table*}[!h]
\centering
\small
\resizebox{\linewidth}{!}{
\begin{tabular}{|p{0.15\linewidth}|p{0.6\linewidth}|p{0.25\linewidth}|}
\hline
\multicolumn{3}{|l|}{\textbf{Dialogue Context}} \\ \hline
\multicolumn{3}{|p{\linewidth}|}{
\textbf{Tutor:} Hi, could you please provide a step-by-step solution for the question below? The question is: The toothpaste in Anne's family's bathroom contains 105 grams. Anne's dad uses 3 grams at each brushing, her mom uses 2 grams, Anne and her brother use 1 gram each. Each member of the family brushes their teeth three times a day. How many days will it take for the toothpaste to run out?

\textbf{Student:} Each family member uses a total of 3+2+1+1 = 7 grams of toothpaste per day. For the whole family, they use 7 x 4 = 28 grams of toothpaste per day. Therefore, the toothpaste will last for 105/28 = 3.75 days. Since they can't use a partial day of toothpaste, the toothpaste will last for 3 days. 3
 
\textbf{Tutor:} Hello, Luca! Can you walk me through your solution?

\textbf{Student:} Sure! I calculated that each family member uses 7 grams of toothpaste per day. So for the whole family, they use 28 grams of toothpaste per day. Then I divided 105 by 28 to get 3.75 days. Since they can't use a partial day of toothpaste, I concluded that the toothpaste will last for 3 days.
} \\ \hline\hline
\textbf{Tutor} & \textbf{Tutor Response} & \textbf{Mistake Identification} \\ 
\hline
\texttt{Gemini} & That's a great start, Luca!  It looks like you've correctly calculated the daily usage, but remember each family member brushes three times a day, so you need to account for that in your total daily usage. & To some extent \\ \hline
\end{tabular}
}
\caption{This case from the test set in the {\em Mistake Identification} track was the most challenging for participating teams. The gold label is "To some extent", as the tutor correctly identified that the student forgot to account for all family members brushing their teeth three times a day. This captures one part of the student's mistake. However, the tutor failed to notice another critical error: the student's incorrect assumption that each family member uses 7 grams of toothpaste. Since only part of the mistake was recognized, the tutor's response is considered a partial identification.}
\label{tab:most_difficult_MI}
\end{table*}

\begin{table*}[!h]
\centering
\small
\resizebox{\linewidth}{!}{
\begin{tabular}{|p{0.2\linewidth}|p{0.8\linewidth}|}
\hline
\textbf{Tutor} & \textbf{Tutor Response} \\ \hline
{\tt Llama-3.1-405B} & Great job on calculating how many beads Caitlin has left after making one bracelet - now, let's think about how many bracelets she can make before she runs out of small beads. \\ \hline
{\tt Llama-3.1-8B} & Now, can you multiply 80 by 2? \\ \hline
\end{tabular}
}
\caption{These responses were the most challenging for participating teams in the {\em Tutor Identification} track -- none of the teams correctly identified the tutor identities. The first response was frequently misclassified as coming from \texttt{Sonnet}, while the second was often labeled as \texttt{Mistral} or \texttt{Expert}.}
\label{tab:most_difficult_TI}
\end{table*}

\end{document}